%% file: platomix3.tex
\documentclass[fleqn,usenatbib]{mnras}

\usepackage{newtxtext} 

\usepackage[T1]{fontenc}
\usepackage{ae,aecompl}

\usepackage{graphicx}	
\usepackage{amsmath}	
\usepackage{amssymb}	

\newcommand{\Msun}{$\mathrm{M}_\odot$}
\newcommand{\Rsun}{$\mathrm{R}_\odot$}
\newcommand{\Ni}{$^{56}$Ni}

\newcommand{\appropto}{\mathrel{\vcenter{
  \offinterlineskip\halign{\hfil$##$\cr
    \propto\cr\noalign{\kern2pt}\sim\cr\noalign{\kern-2pt}}}}}

\title[Ni-mixing in SNe~IIP]{The role of radioactive nickel in shaping the plateau phase of Type\,II supernovae}

\author[A. Kozyreva et al.]{
Alexandra~Kozyreva,$^{1,2}$\thanks{E-mail: sasha@mpa-garching.mpg.de}
Ehud Nakar,$^{2}$
Roni Waldman$^{3}$
\\
$^{1}$Max-Planck-Institut f\"ur Astrophysik, Garching bei M\"unchen, 85748, Germany\\
$^{2}$The Sackler School of Physics and Astronomy, Tel Aviv University, Tel Aviv, 6997212, Israel\\
$^{3}$Racah Institute of Physics, The Hebrew University, Jerusalem 91904, Israel\\
}

\date{Accepted XXX. Received YYY; in original form ZZZ}

\pubyear{2018}

\begin{document}
\label{firstpage}
\pagerange{\pageref{firstpage}--\pageref{lastpage}}
\maketitle

\begin{abstract} 
In the present study, we
systematically explore the effect of the radioactive $^{56}$Ni and its
mixing properties in the ejecta on the plateau of Type\,IIP supernovae (SNe).  
We evaluate the importance of $^{56}$Ni in shaping light curves of
SNe\,IIP by simulating light curves for two red supergiant models using different amounts of $^{56}$Ni and with
different types of mixing: uniform distribution of $^{56}$Ni out to different fractions of the envelope and ``boxcar''
distribution of $^{56}$Ni.  We
find, similarly to previous studies, that \Ni{} extends duration of the
plateau.  We find a formula to estimate the extension based on the observed
bolometric light curves and show that for most SNe\,IIP $^{56}$Ni extends
the plateau by about 20\,\%{}.  Another effect of \Ni{} consists in reduction
of the plateau decline rate, i.e. \Ni{} presented in the ejecta flattens the
plateau.  Our simulations
suggest that for typical SNe\,IIP it can reduce the decline rate by about
1\,mag/100\,day. We find that for the contribution of \Ni{} seen in most SNe our
simulated bolometric light curves resemble observed ones for various types of \Ni{} mixing. We thereby cannot
determine the level of \Ni{} mixing in these SNe based on the light curve alone. 
However, for SN\,2009ib  we find that 
only a model where \Ni{} is mixed significantly throughout most of the hydrogen envelope 
is consistent with the observed light curve.
Our light curves are available via link \url{https://wwwmpa.mpa-garching.mpg.de/ccsnarchive/data/Kozyreva2018/}.

\end{abstract}




\section[Introduction]{Introduction} 
\label{sect:intro}

Type II supernovae (SNe\,II), i.e. those supernovae (SNe) which display
strong hydrogen lines in the spectra
at the time of discovery, are the most common explosions in the
volume-limited sample \citep{2009ARA&A..47...63S,2011MNRAS.412.1522S,2011MNRAS.412.1473L}. 
Among them, Type~IIP SNe remain bright during about 100 days showing the
so-called plateau phase. They contribute 
50\% to 80\% to all core-collapse explosions. Roughly 75\% of all stellar
explosions are core-collapse SNe \citep{2003ApJ...586....1M,2004ApJ...613..189D,2007MNRAS.377.1229M}.
CCSNe originate from explosions of massive stars, i.e. stars with
initial masses above 8~\Msun{} and below 100~\Msun{}.
Progenitors for SNe\,II are stars which retain hydrogen envelopes by the
time of iron-core collapse, among which extended red supergiants produce
SNe\,IIP \citep{1960SvA.....4..355S,1971Ap&SS..10....3G,2009ARA&A..47...63S}. 

\begin{table*}
\centering
\caption{Key characteristics of the input models.}
\label{table:models}
\begin{center}
\begin{tabular}{|l|c|c|c|c|c|c|c|c|c|c|c|c|c|}
\hline
model & Radius
&M$_\mathrm{tot}$&M$_\mathrm{H}$&M$_\mathrm{He}$&M$_\mathrm{C}$&M$_\mathrm{O}$&M$_\mathrm{Ne}$&M$_\mathrm{Ni}$& \multicolumn{4}{c}{mixed in} & E$_\mathrm{expl}$\\
      & [\Rsun{}]& [\Msun{}]    &[\Msun{}]      &[\Msun{}]     &[\Msun{}]     &[\Msun{}]      &[\Msun{}]      &[\Msun{}] & \multicolumn{4}{c}{[fraction of ejecta]} & [foe($\equiv10^{\,51}$\,erg)]\\
\hline
m12 & 496    &11.25           & 5.4   & 3.8&0.09& 0.6&0.04&  0    & centre       & \multicolumn{3}{c}{uniform} & 0.4\\
    &        &                &       &    &    &    &    &  0.011& 0.22~\Msun{} & 1/3  &  2/3   & 1           & 0.9\\
    &        &                &       &    &    &    &    &  0.025&              &      &        &             & 1.35\\
    &        &                &       &    &    &    &    &  0.045&              &      &        &             & \\
    &        &                &       &    &    &    &    &  0.065&              &      &        &             & \\
    &        &                &       &    &    &    &    &  0.137&              &      &        &             & \\
\hline
m15 & 631    &13.4            & 6.0   & 4.3&0.17& 0.2&0.24&  0    & centre       & \multicolumn{2}{c}{boxcar} & & 0.53\\
    &        &                &       &    &    &    &    &0.028  &0.4~\Msun{}   &0.16  &  0.31                & & 1.1\\
    &        &                &       &    &    &    &    &0.056  &              & \multicolumn{2}{c}{uniform}& & 1.53\\
    &        &                &       &    &    &    &    &0.113  &              &0.46  &  0.88                & & \\
\hline
\end{tabular}
\end{center}
\end{table*}

\begin{figure*}
\centering
\includegraphics[width=0.5\textwidth]{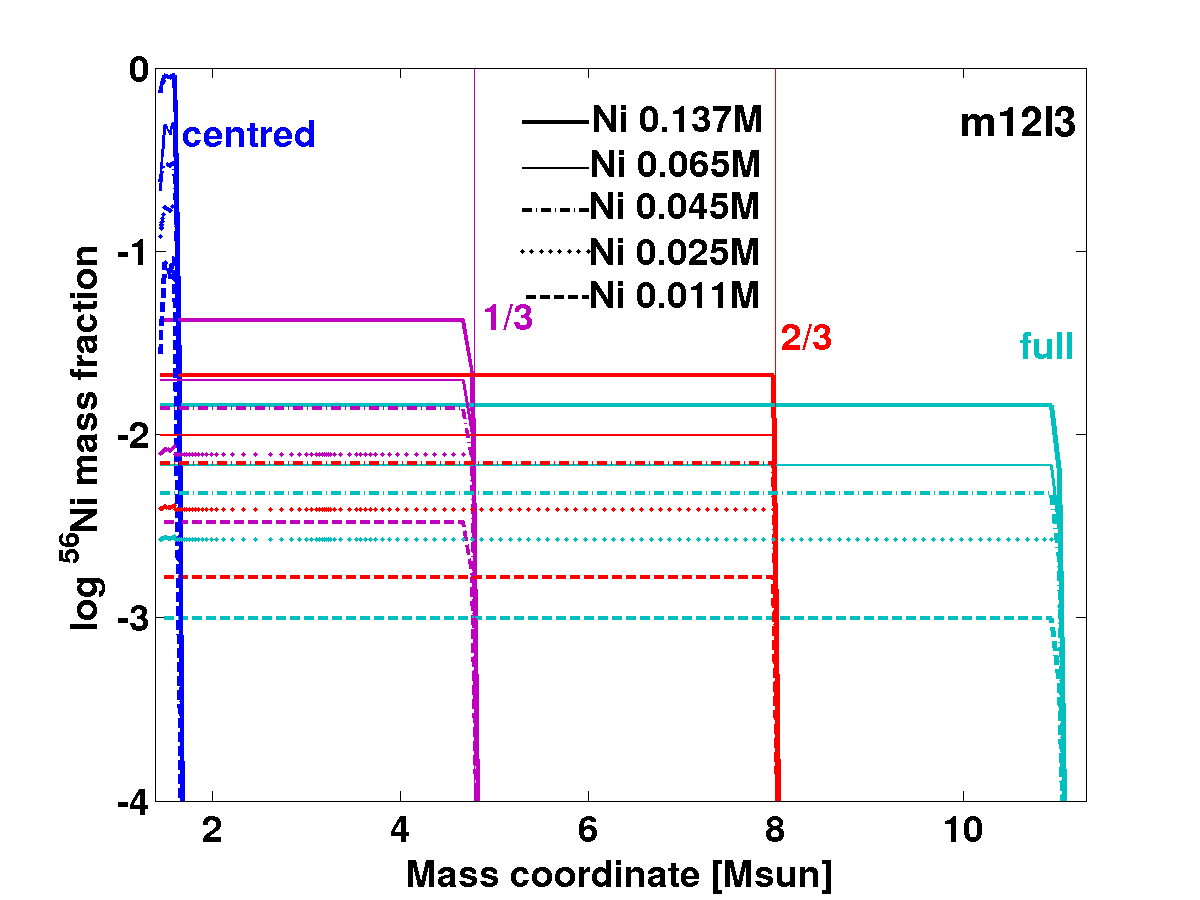}~
\includegraphics[width=0.5\textwidth]{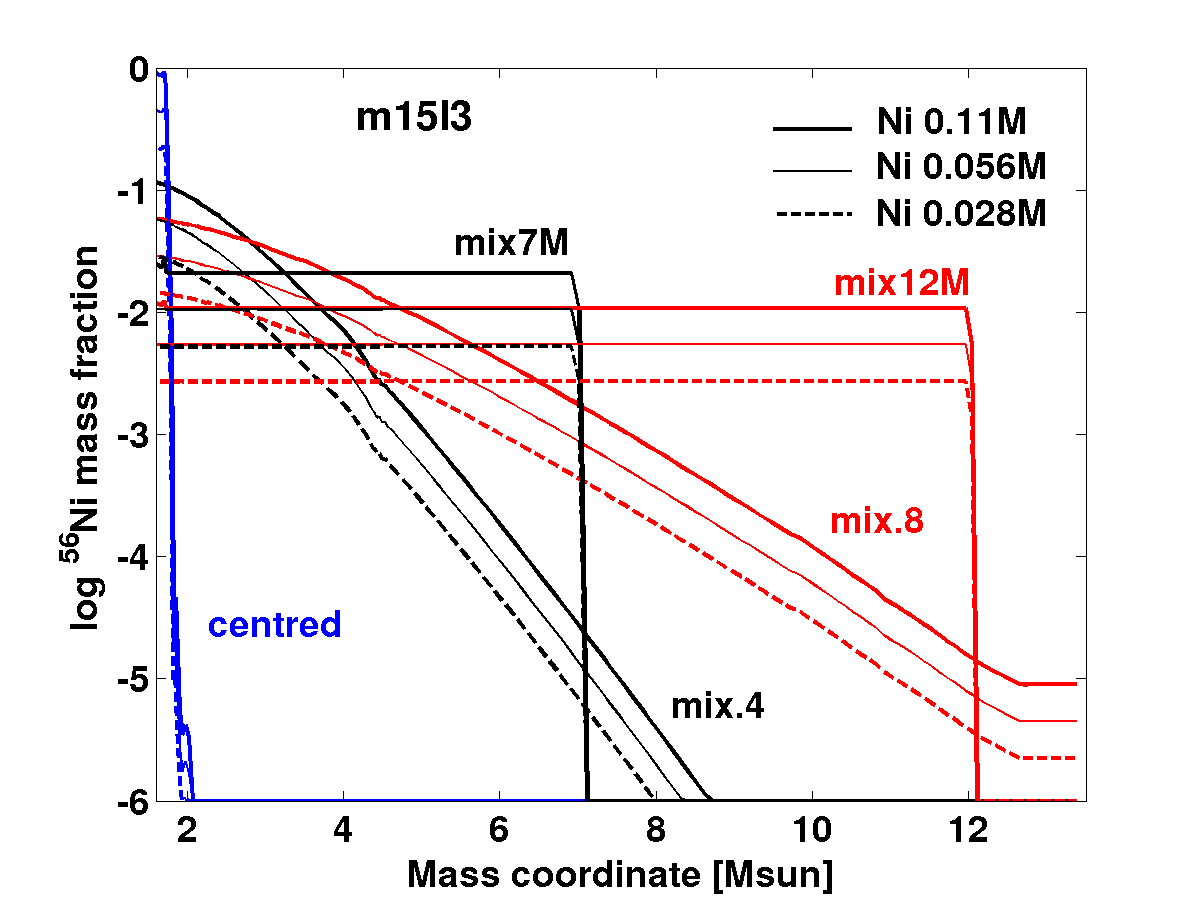}
\caption{Distribution of \,\,$^{56}$Ni in the ejecta of the model m12 (left)
and of the model m15 (right). See explanation in the text.}
\label{figure:Nidistr}
\end{figure*}

A given progenitor with a particular radius, mass, density, chemical
structure, and explosion energy defines a unique light curve. However, the common task is to
solve the reverse problem, i.e.
determine the progenitor parameters from observational signatures of the
SN explosion. Early studies estimated the progenitor and explosion parameters 
from the SN\,IIP light curve and photospheric velocity observations taking into account that 
the plateau phase is supported mostly by the thermal energy deposited by the shock
wave which unbinds the progenitor envelope and that its evolution is dictated by the 
cooling and recombination wave receding
through the expanding envelope \citep{1971Ap&SS..10....3G,1976Ap&SS..44..409G,1977ApJS...33..515F,1985SvAL...11..145L,1993ApJ...414..712P}.
These studies provide formulae that relate light curve properties, mostly the plateau luminosity 
and duration and the photospheric velocity, to the most general progenitor properties (mass and 
radius) and the explosion energy. This is done using  models with different levels of approximations 
of the the recombination wave that crosses the hydrogen envelope, neglecting 
contribution from freshly made radioactive $^{56}$Ni. However, more recent studies 
have shown that the energy input from radioactive decay of nickel \Ni{} and cobalt $^{56}$Co strongly affects 
the behavior of the cooling wave and the resulting observations
in SNe~IIP \citep[][and others]{2004ApJ...617.1233Y,2009ApJ...703.2205K,2011ApJ...729...61B}.
As a result, estimates which ignore the contribution of $^{56}$Ni can be highly inaccurate. 

The energy deposited by the radioactive decay of $^{56}$Ni starts affecting
the observed emission as soon as the recombination wave encounters 
Ni-generated photons diffusing through the inner ejecta.
Once that happens $^{56}$Ni energy deposition tends to increase the luminosity
compared to the emission if no $^{56}$Ni was present.  Since $^{56}$Ni
contribution is more dominant at later times, it has two major effects. 
First, it delays the propagation of the recombination wave, thereby
extending the plateau duration \citep[e.g.,][]{2009ApJ...703.2205K}. 
Second, it reduces luminosity decline rate, making the plateau
``flatter'' \citep[e.g.,][]{2011ApJ...729...61B}.  The exact effect depends on
the total abundance of \Ni{} and on its mixing throughout the envelope.

Recently, \citet{2016ApJ...823..127N} introduced an observable that measures the 
importance of $^{56}$Ni in the light curve of Type II SNe.  They also analysed 24 
observed SNe~IIP and evaluated the importance
of $^{56}$Ni heating for the plateau phase. They concluded that $^{56}$Ni
contributes to most SN~IIP plateaus in their study and plays an important 
role where the effect consists of both an extension and a flattening of the 
plateau. In Section~\ref{sect:eta} we give a brief description of the observable 
introduced by \citet{2016ApJ...823..127N} and of the results of their analysis.

The goal of this paper is to study numerically the effect of $^{56}$Ni on the light curve of type II SNe, 
and especially how the signature of $^{56}$Ni depends on its mixing through the envelope. In order to do 
that we carry out simulations of two red supergiant
progenitors and their explosions (Section~\ref{sect:models}). We explore the effect of Ni-heating during
photospheric phase, by varying the amount of $^{56}$Ni and its mixing as well as the explosion
energy (Section~\ref{subsect:mix}), 
and analyse the obtained light curves in the context of observations using the measures introduced
in \citet{2016ApJ...823..127N} (Section\,\ref{sect:results}). We summarise the conclusions of our study in Section\,\ref{sect:conclusions}.

\section[Eta]{Observable that measures the importance of \Ni{}}
\label{sect:eta}

The main difficulty in measuring the effect of \Ni{} is that radiative
transfer couples the energy deposited by the \Ni{} to the energy deposited by
the SN shock (e.g., by affecting the ionization and thereby the opacity) in
a way that the instantaneous luminosity cannot be separated to the
contribution of each component.  However, \citet{2016ApJ...823..127N} have
shown that there are integrated observable quantities where this separation
is possible.  They generalized the result of \citet{2013arXiv1301.6766K}
that have shown that the integral of the time weighted bolometric luminosity
is a highly accurate measure of the integral over the time weighted energy
deposition.  Now, since the amount of $^{56}$Ni in Type II SNe can be
measured quite accurately from their nebular phase, one can separate the
time weighted energy deposition to that of the initial energy deposited by
the shock and the additional energy deposited by radioactive decay. 
\citet{2016ApJ...823..127N}, therefore, defined the observable:

\begin{equation}
\eta_\mathrm{Ni} = \frac{\int_0^{t_\mathrm{Ni}} t \, Q_\mathrm{Ni56} \, dt}{\int_0^{t_\mathrm{Ni}} t
\,(L_\mathrm{bol}-Q_\mathrm{Ni56}) \, dt} \, ,
\label{eta}
\end{equation}

where $t$ is the time since the explosion, $L_\mathrm{bol}(t)$ is the
instantaneous bolometric luminosity and $Q_\mathrm{Ni56}(t)$ is the
instantaneous deposition of energy by radioactive decay.  $t_\mathrm{Ni}$ is the
time that the photospheric phase ends, which is marked by the end of
luminosity drop at the end of the plateau and the beginning of the
the $^{56}$Co tail. 

The time weights in the integrals account for the adiabatic losses of
the radiation between the time that the energy is deposited and the time
that it is released, making this observable to be one of the few that are
insensitive to the unknown details of the radiation transfer through the
envelope.  The numerator is insensitive to the radiation deposited by the SN
shock.  it measures the integrated time weighted luminosity that we would
have seen if all the emission were powered by \Ni{} (as, for example, in Type
I SNe) and it is roughly proportional to $M_\mathrm{Ni56}$.  The denominator
is insensitive to the presence of \Ni{} and it  measures the integrated time
weighted luminosity that we would have seen if there were no \Ni{}. 
\citet{2016arXiv160202774S} studied the physical meaning of the denominator,
which they denote as ET.  They show that it depends on the progenitor
structure as well as the explosion energy and that for red-supergiants that
retain most of their H envelope it can be roughly approximated as $ET
\appropto E^{1/2} M_{ej}^{1/2} R_0$, where $E$ is the kinetic energy at
infinity (defined as the explosion energy), $M_{ej}$ is the ejecta mass and
$R_0$ is the progenitor radius.

$\eta_\mathrm{Ni}$ is a measure of the importance of \Ni{} in shaping the emission that we see. 
If $\eta_\mathrm{Ni} \ll 1$ then \Ni{} is unimportant and there is a little difference between 
the observed light curve and the one that would have been observed if there were no \Ni{}.  
If $\eta_\mathrm{Ni} \gg 1$ it implies that most of the observed emission is generated by 
\Ni{} (this is the case in type I SNe). \citet{2016ApJ...823..127N} analysed 24 type II SNe 
with a good bolometric (or pseudo-bolometric) light curves and calculated the value of 
$\eta_\mathrm{Ni}$ for these SNe. They find that for all SNe except one, $\eta_\mathrm{Ni}$ 
falls within the range $0.1-0.7$. SN\,2009ib is an exception
which has $\eta_\mathrm{Ni}=2.6$. 

\Ni{} is expected to affect the decline rate during the plateau as well, and indeed \citet{2016ApJ...823..127N} 
found that $\eta_\mathrm{Ni}$ is anti-correlated with the decline rate. \Ni{} 
is also supposed to extend the duration of the plateau. In the following section we examine these 
expectations using numerical simulations.

\section[Input models]{Input models} 
\label{sect:models}

For our analysis we computed two hydrogen-rich red supergiant models m12
and m15 with initial masses of 12~\Msun{} and 15~\Msun{}, correspondingly
(see Table~\ref{table:models}).
The models are at solar metallicity and non-rotating. The mixing-length parameter is chosen
equal to 3. 
The main property of the models is the presence of a hydrogen-rich (total
hydrogen mass 5.4~\Msun{} and 6~\Msun{}) extended envelope (496~\Rsun{} and 631~\Rsun{}).
We apply the following method. Firstly,
the stellar evolution from zero-age main sequence until the formation of an
iron core was computed with 
\verb|MESA|\footnote{Modules for Experiments in Stellar Astrophysics
\url{http://mesa.sourceforge.net/}
\citep{2011ApJS..192....3P,2013ApJS..208....4P,2015ApJS..220...15P}.}. 
Secondly, the models were blown up with \verb|V1D|.
Explosion is created by means of a piston, which is set at a
Lagrangian mass of choice, given initial velocity equal to the escape
velocity, and then allowed to free-fall. 
Thirdly, were mapped into the radiation hydrodynamics code \verb|STELLA| to
follow the post-explosion evolution.
\verb|V1D| is a one-dimensional hydrodynamics version of the code
\verb|Vulcan| \citep{1993ApJ...412..634L}. 
\verb|V1D| solves the equations of motion using explicit Lagrangian hydrodynamics. The
radiative-transport in \verb|V1D| is solved under the assumption of LTE and diffusion approximation
for radiative transfer. The opacities in \verb|V1D| are computed based on the opacity
routines of \verb|CMFGEN| 
\citep{2010MNRAS.405.2141D,2010MNRAS.405.2113D,2015MNRAS.449.4304D}. 
Our main light curve simulations are carried out with \verb|STELLA| which is a one-dimensional hydrodynamics
code which solves radiative transfer equations in hundred frequency bins in
momentum approximation \citep{1998ApJ...496..454B,2006A&A...453..229B}.
Additionally, we carried out radiative transfer simulations with the
multi-group extension to \verb|V1D| and compare to the main results computed
with \verb|STELLA|.

The explosion of the m12 and m15 models by \verb|V1D| was done using default explosion energy of 0.9~foe and 1.1~foe,
correspondingly. To vary explosion energy, we modified velocity profile of the shocked material 
via multiplying by a certain factor while mapping the models into \verb|STELLA|. The explosion 
energy of the models we run is defined as the kinetic energy at infinity and it is: 0.4~foe, 0.9~foe, and 1.35~foe
for m12, and 0.53~foe, 1.1~foe, and 1.53~foe for m15.

\subsection[Nickel mixing setup]{Nickel mixing setup} 
\label{subsect:mix}

In Figure~\ref{figure:Nidistr}, we demonstrate the input profiles for \,\,$^{56}$Ni
distribution in the ejecta.
The values of $^{56}$Ni mass that we set are, 0.011~\Msun{}, 0.025~\Msun{},
0.045~\Msun{}, 0.065~\Msun{}, 0.14~\Msun{}, and no nickel for m12 (Fig.~\ref{figure:Nidistr} left), and
0.028~\Msun{}, 0.056~\Msun{}, 0.11~\Msun{}, and no nickel for m15 (Fig.~\ref{figure:Nidistr} right).
By default $^{56}$Ni is concentrated to the inner 0.22~\Msun{} and
0.4~\Msun{} in the model m12 and the model m15, correspondingly.
Default distribution of \Ni{} comes from \verb|V1D| 
simulations of the piston-driven explosion. Nucleosynthesis in \verb|V1D| 
is done with the implemented nuclear network which includes 54 isotopes.
Reaction rates are as given in the non-smoker database
\url{https://nucastro.org/nonsmoker.html}.

Throughout
the paper we call this kind of unmixed distribution as ``centrally located''
or ``centrally concentrated'' $^{56}$Ni.
We apply two kinds of mixing. For the model m12, we uniformly spread
radioactive nickel in 1/3, 2/3 and in the entire (so-called ``full'') ejecta mass. 
For uniform mixing, \Ni{} is set as shown in Figure~\ref{figure:Nidistr} while mass fraction of the rest species are recalibrated in each Lagrangian zone to keep the sum of mass fraction equal unity.
For the model m15, we applied both uniform and so-called
``boxcar'' mixing. For boxcar mixing, we loop over all zones of the model.
For each zone, with Lagrangian mass $m$, we uniformly mix the composition in all 
zones between $m$ and $m+dm$, where $dm$ is the boxcar parameter, e.g. 0.4, 0.8, 
in \Msun{} units. We repeat the above procedure a total of four times.

The ``boxcar'' method is supposed to 
imitate mixing of chemical elements taking place during the earlier phase of expansion
in core-collapse explosions.
``Mix.4'' means $^{56}$Ni distribution
in which we applied ``boxcar''-mixing with the ``boxcar''-parameter 0.4. ``Mix.8''
stands for mixing with the ``boxcar''-parameter 0.8. In fact, 95\,\% of \Ni{}
is located in the inner 3.5~\Msun{} and 5.3~\Msun{} for ``mix.4'' and
``mix.8'', correspondingly, i.e. in 0.16 and 0.31 of the ejecta.
``Mix7M'' and ``Mix12M'' mean uniform mixing of $^{56}$Ni in 7~\Msun{} (46\,\%) and 12~\Msun{} (88\,\%)
of the expanding ejecta, correspondingly.

In total, each evolutionary model has 6 and 4 values for mass of $^{56}$Ni, 3
and 5 kinds of mixing, for m12 and m15 respectively, and 3 values for explosion energy.

\input{results3}
\input{conclusion3}

\section*{Acknowledgments}
AK and EN are supported by ERC grant No.\,279368
(``The Gamma Ray Burst -- Supernova Connection and shock breakout physics'')
and partially by the I-Core center of excellence of the CHE-ISF.
The \verb|STELLA| simulations were carried out on the DIRAC Complexity system (grants ST/K000373/1 and ST/M006948/1),
operated by the University of
Leicester IT Services, which forms part of the STFC DiRAC HPC Facility
(\url{www.dirac.ac.uk}). AK thank Sergey Blinnikov and Viktor Utrobin 
for fruitful and useful discussions.

\addcontentsline{toc}{section}{Acknowledgments}

\bibliographystyle{mnras}
\bibliography{references}
\input{append}








%


\bsp	
\label{lastpage}
\end{document}

%% file: results3.tex

\section[Results]{Results} 
\label{sect:results}

\subsection[Light curves]{Light curves}
\label{subsect:lc}

\begin{figure*}
\centering
\includegraphics[width=0.5\textwidth]{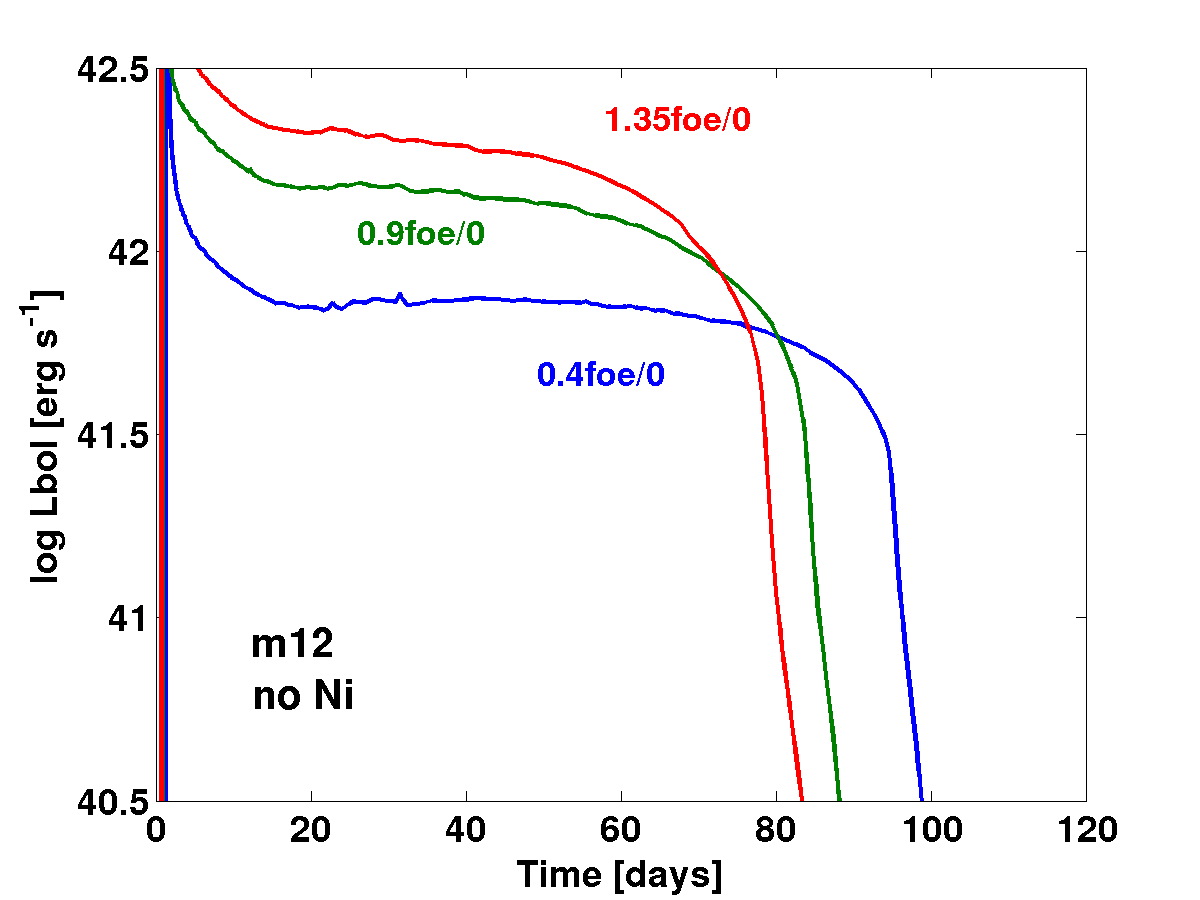}~
\includegraphics[width=0.5\textwidth]{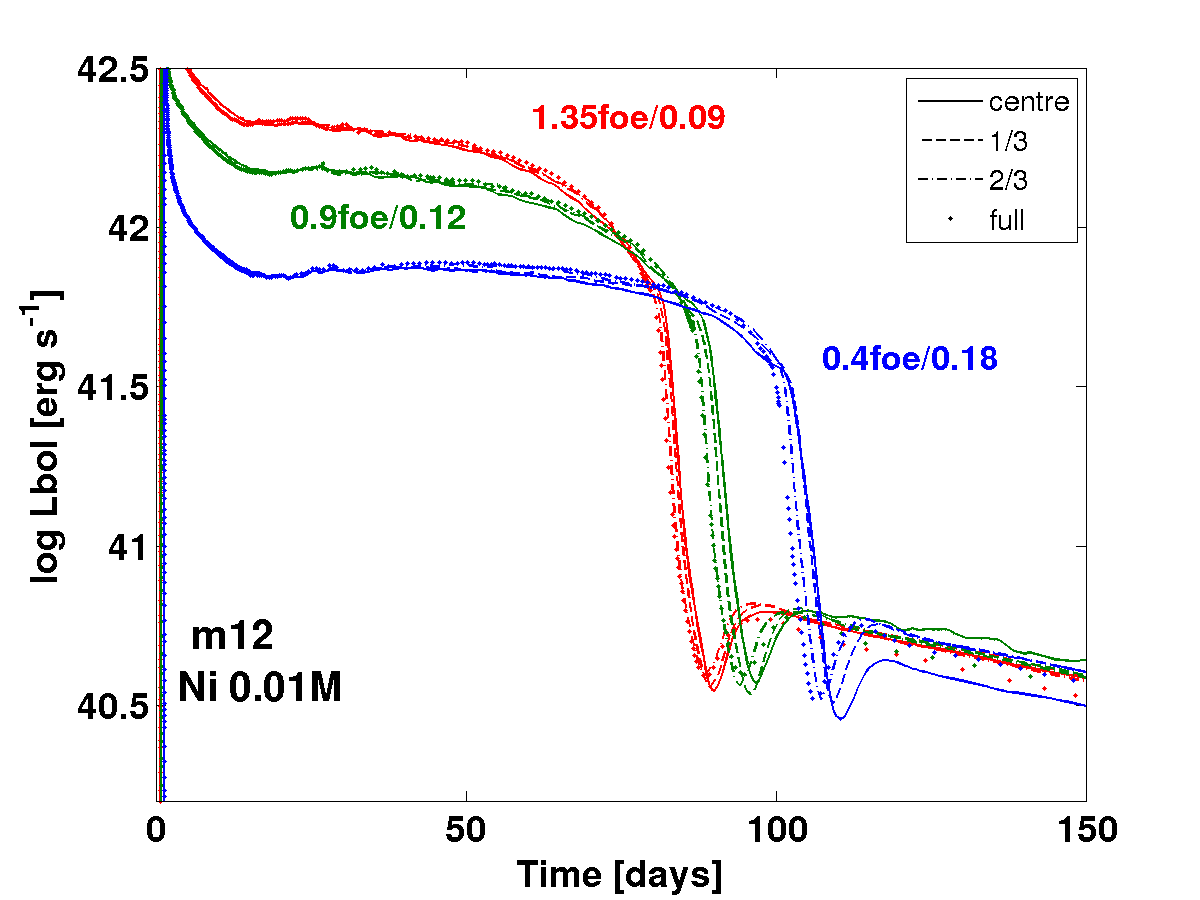}\\
\includegraphics[width=0.5\textwidth]{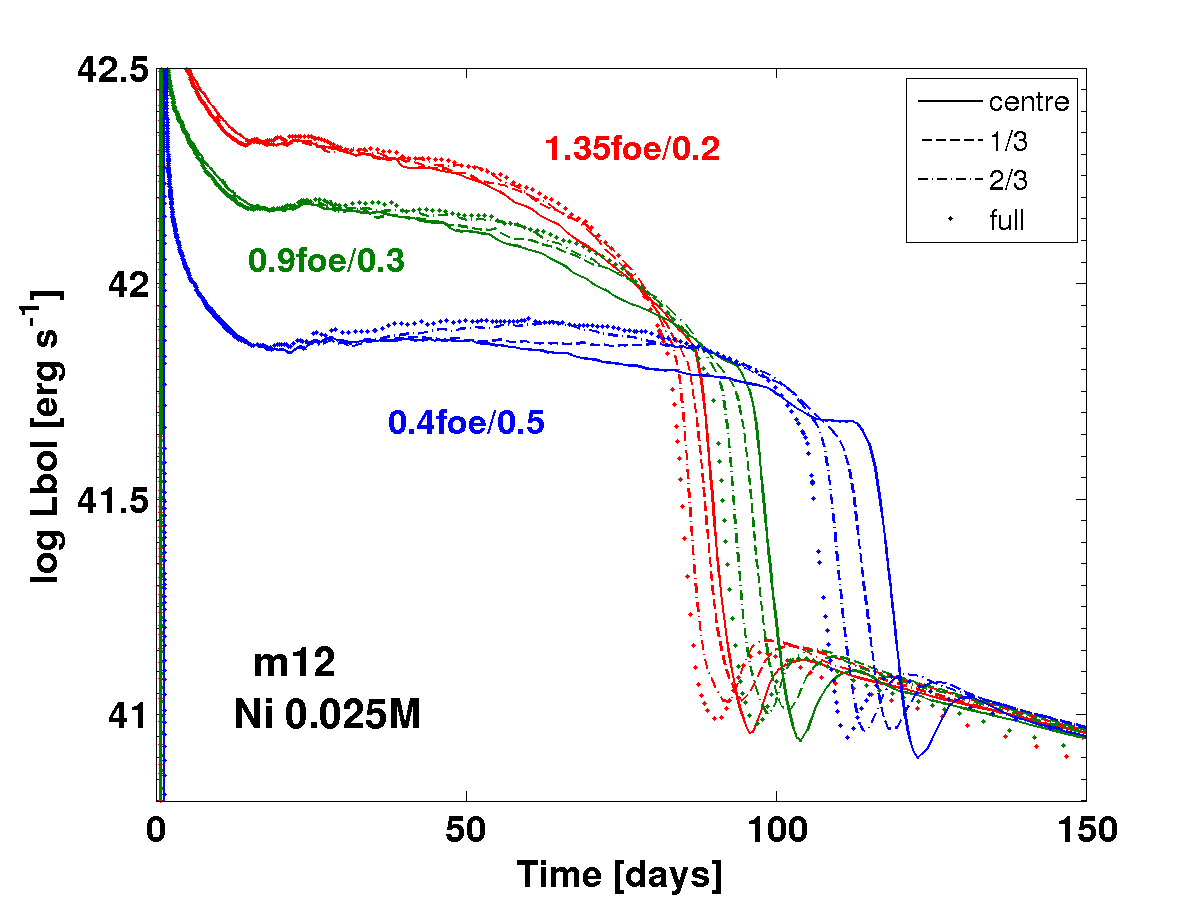}~
\includegraphics[width=0.5\textwidth]{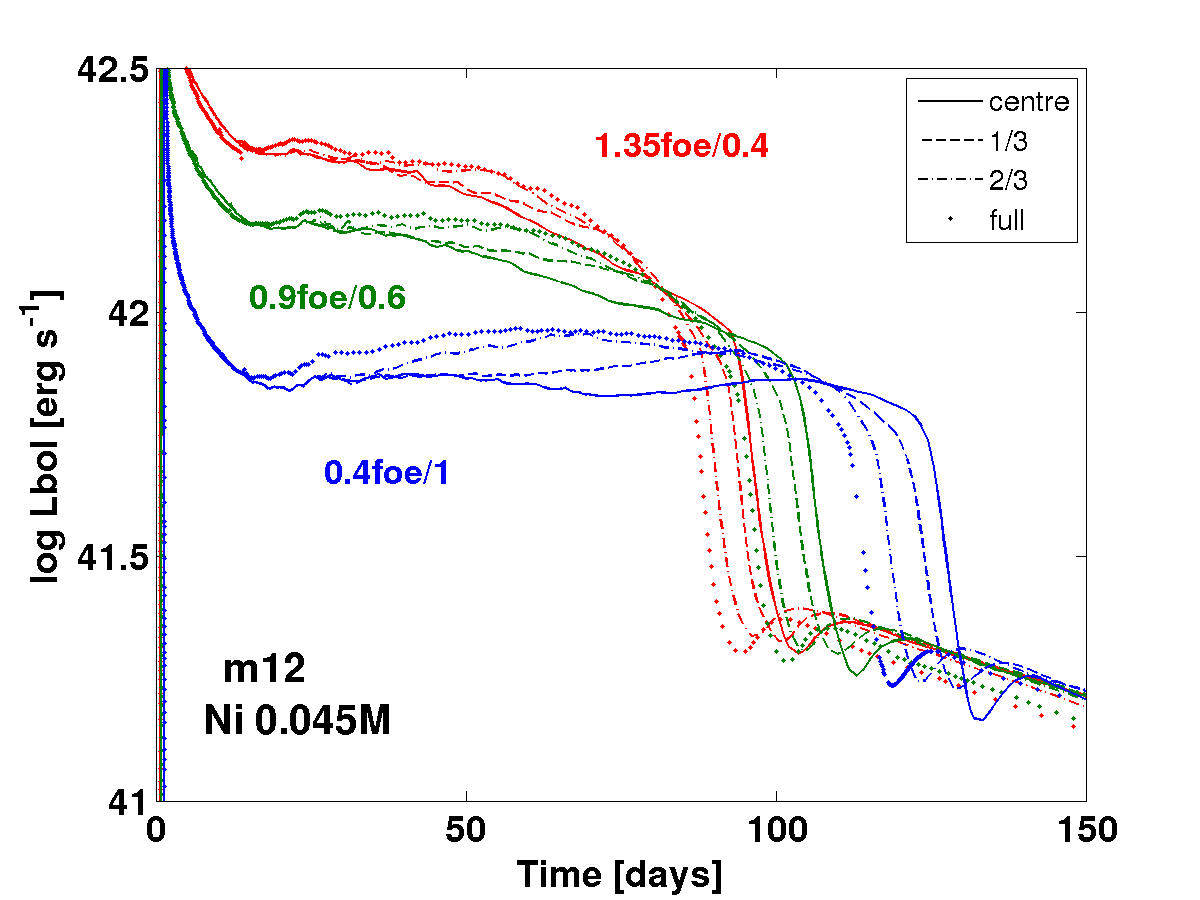}\\
\includegraphics[width=0.5\textwidth]{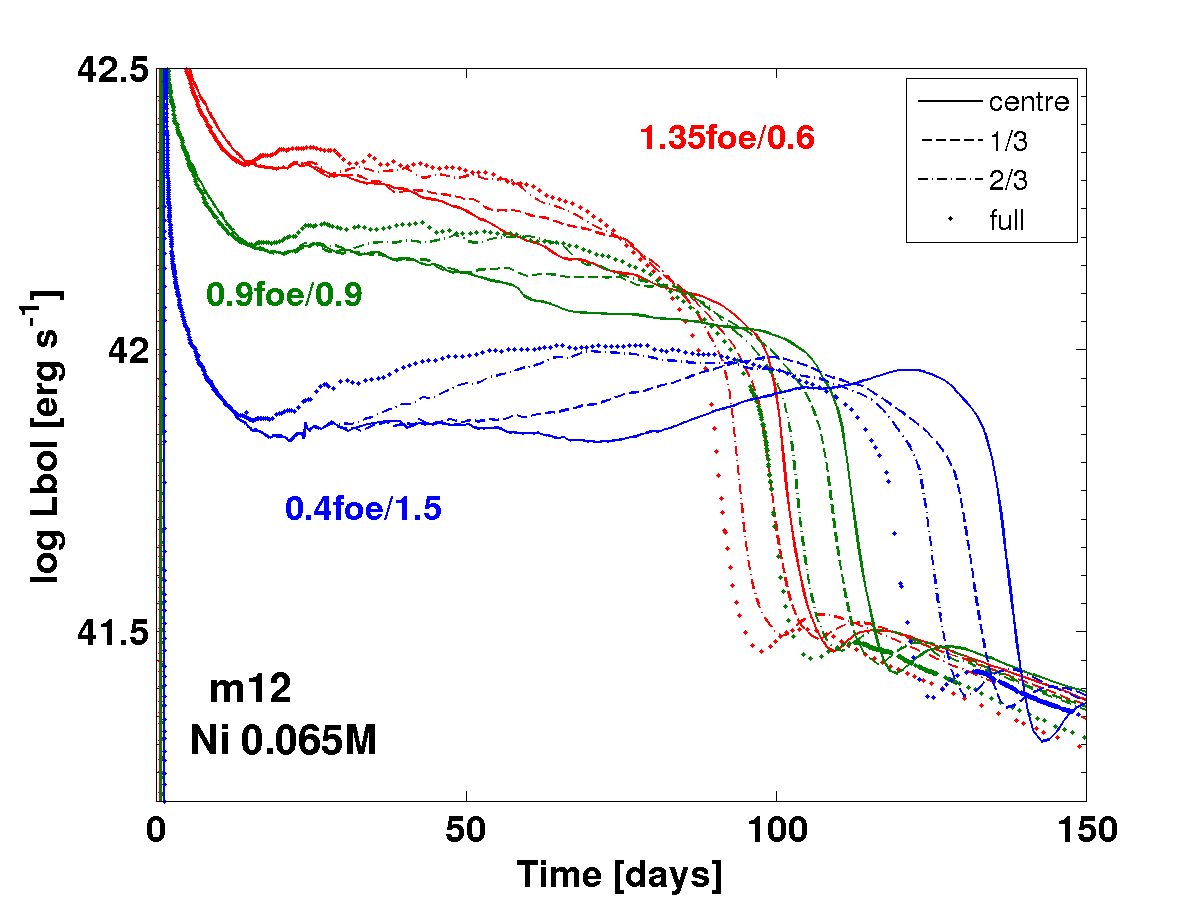}~
\includegraphics[width=0.5\textwidth]{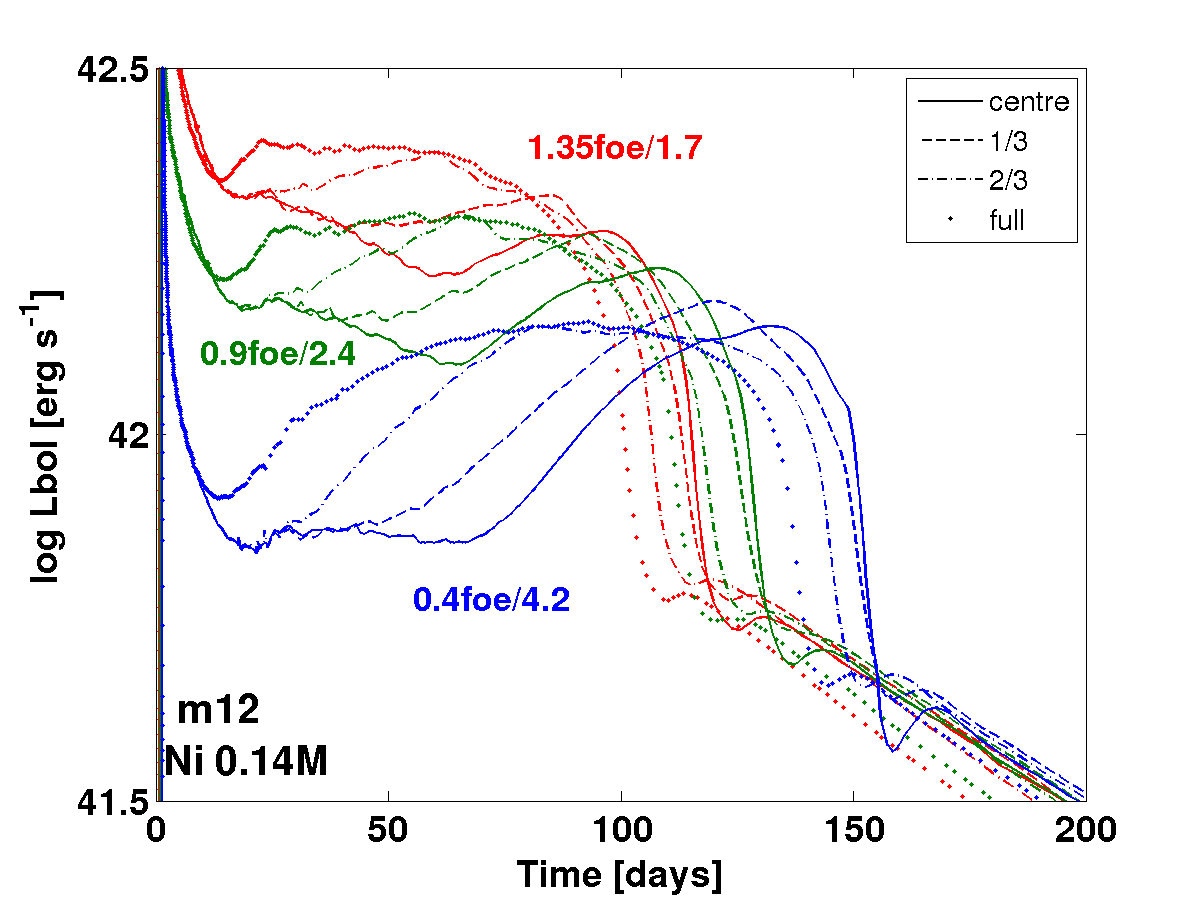}
\caption{Bolometric light curves for the set of m15 models with different $^{56}$Ni
masses, different mixing of $^{56}$Ni, different explosion energies. See
Table~\ref{table:models} for details. The labels indicate the explosion energy and $\eta_\mathrm{Ni}$.}
\label{figure:m12lbol}
\end{figure*}

\begin{figure*}
\centering
\includegraphics[width=0.5\textwidth]{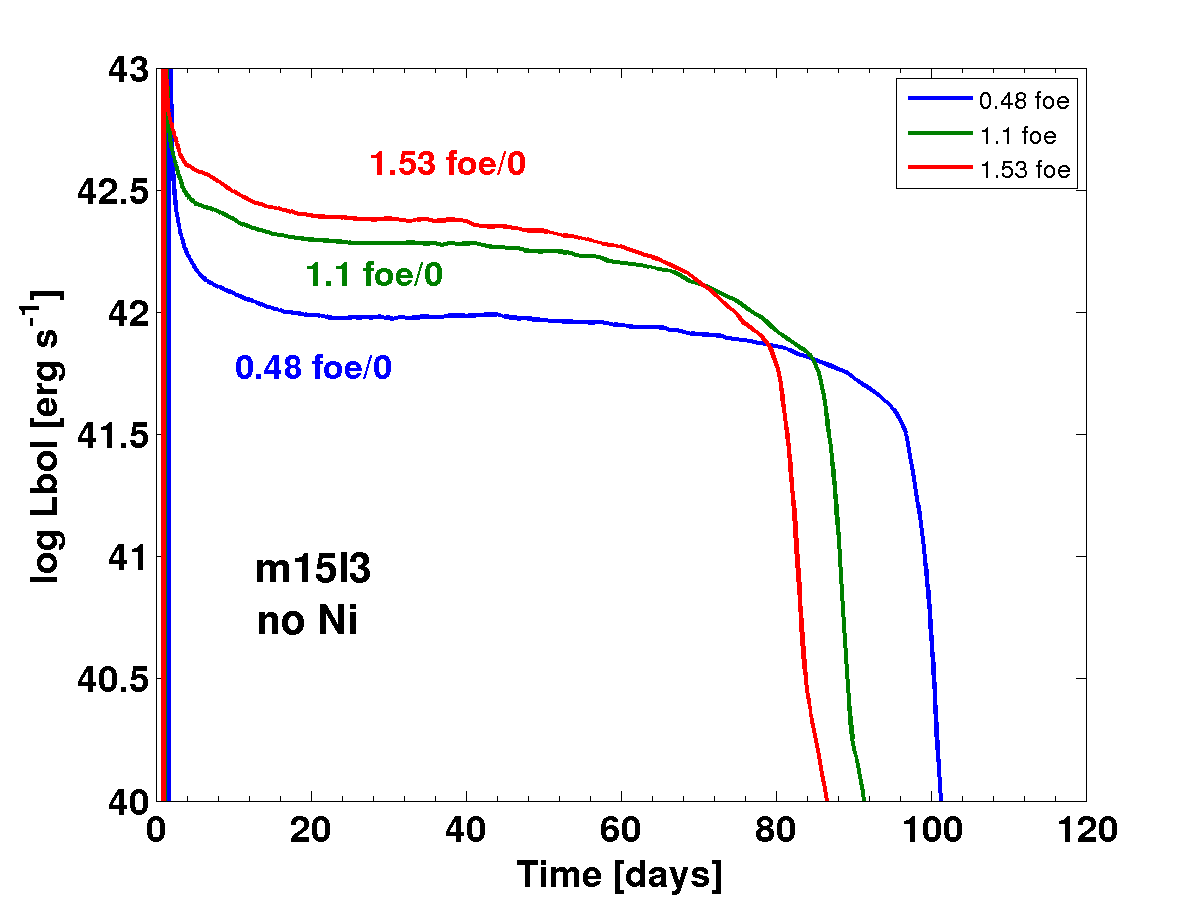}~
\includegraphics[width=0.5\textwidth]{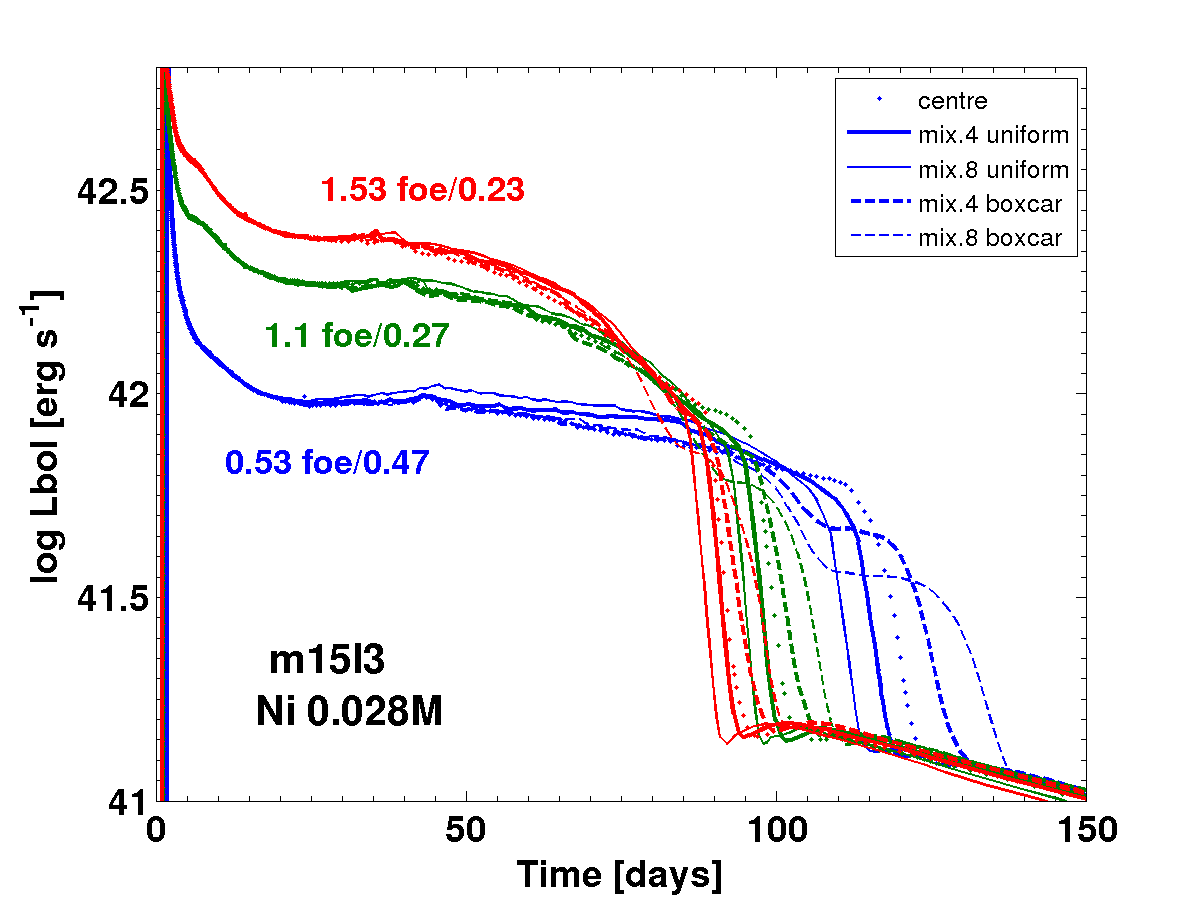}\\
\includegraphics[width=0.5\textwidth]{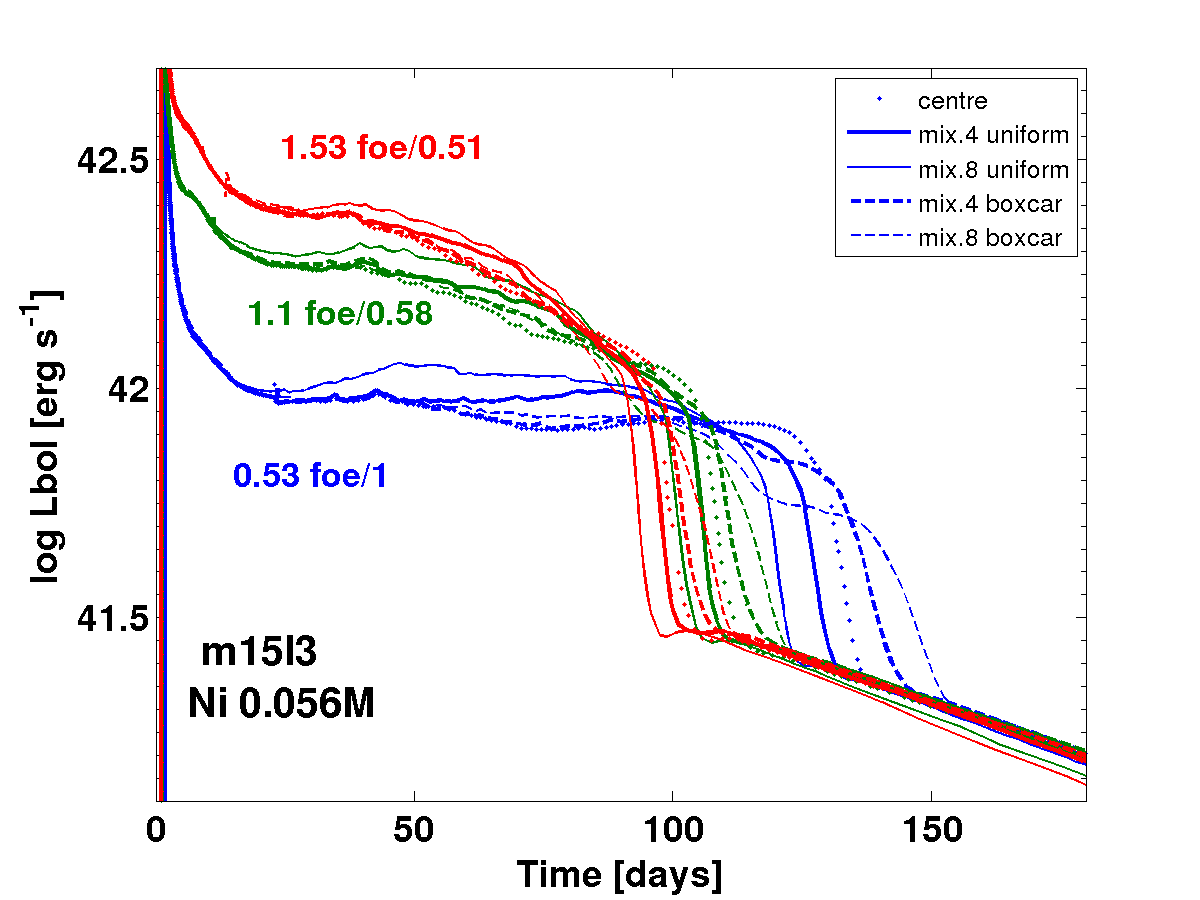}~
\includegraphics[width=0.5\textwidth]{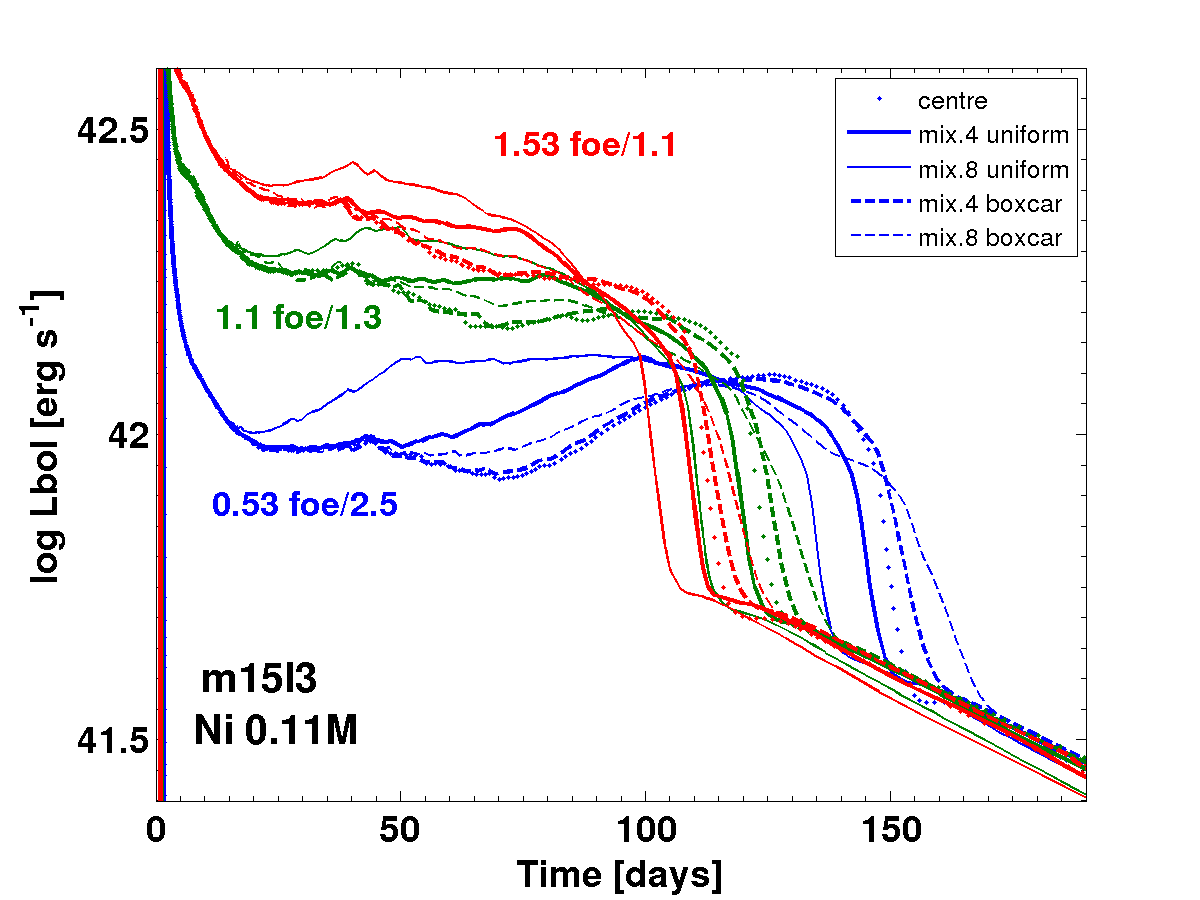}
\caption{Bolometric light curves for the set of m15 models.}
\label{figure:m15lbol}
\end{figure*}

\begin{figure*}
\centering
\includegraphics[width=0.5\textwidth]{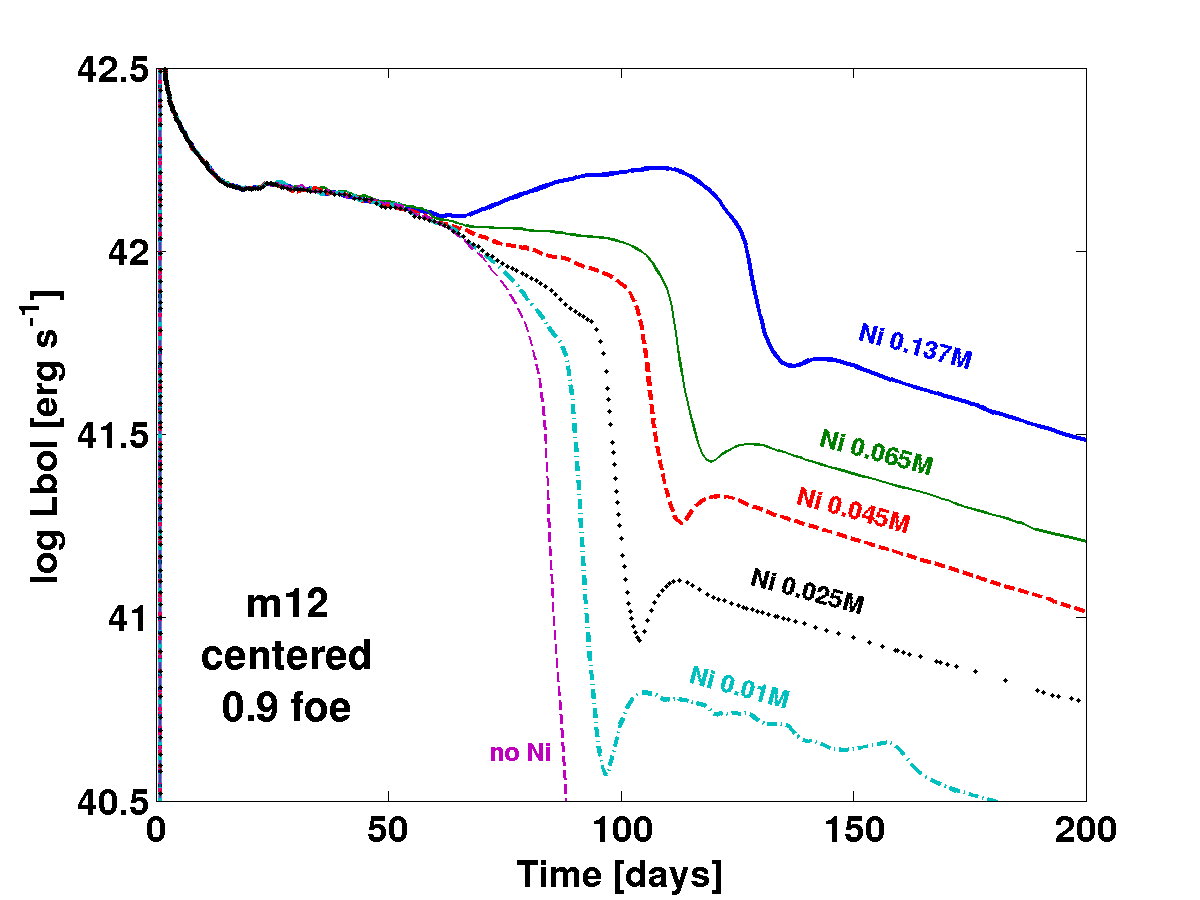}~
\includegraphics[width=0.5\textwidth]{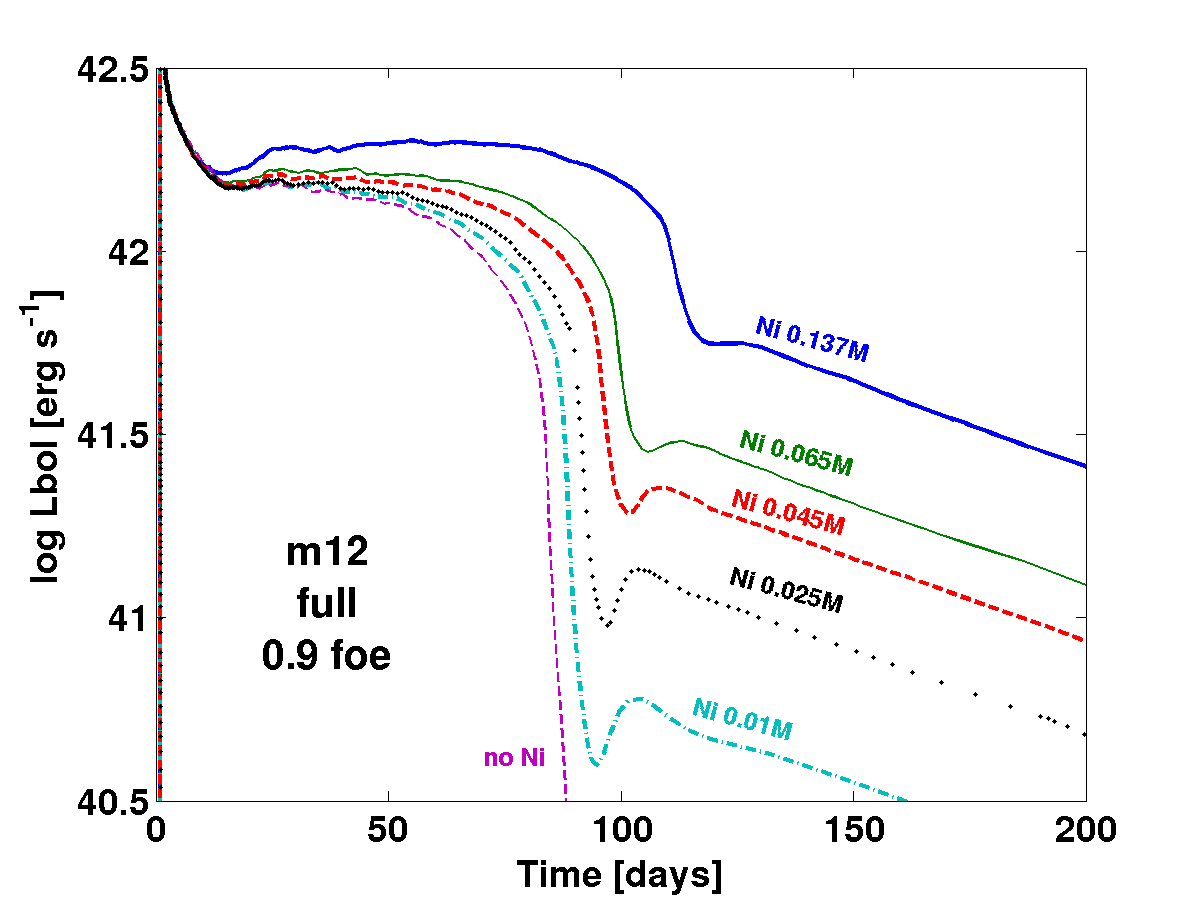}\\
\includegraphics[width=0.5\textwidth]{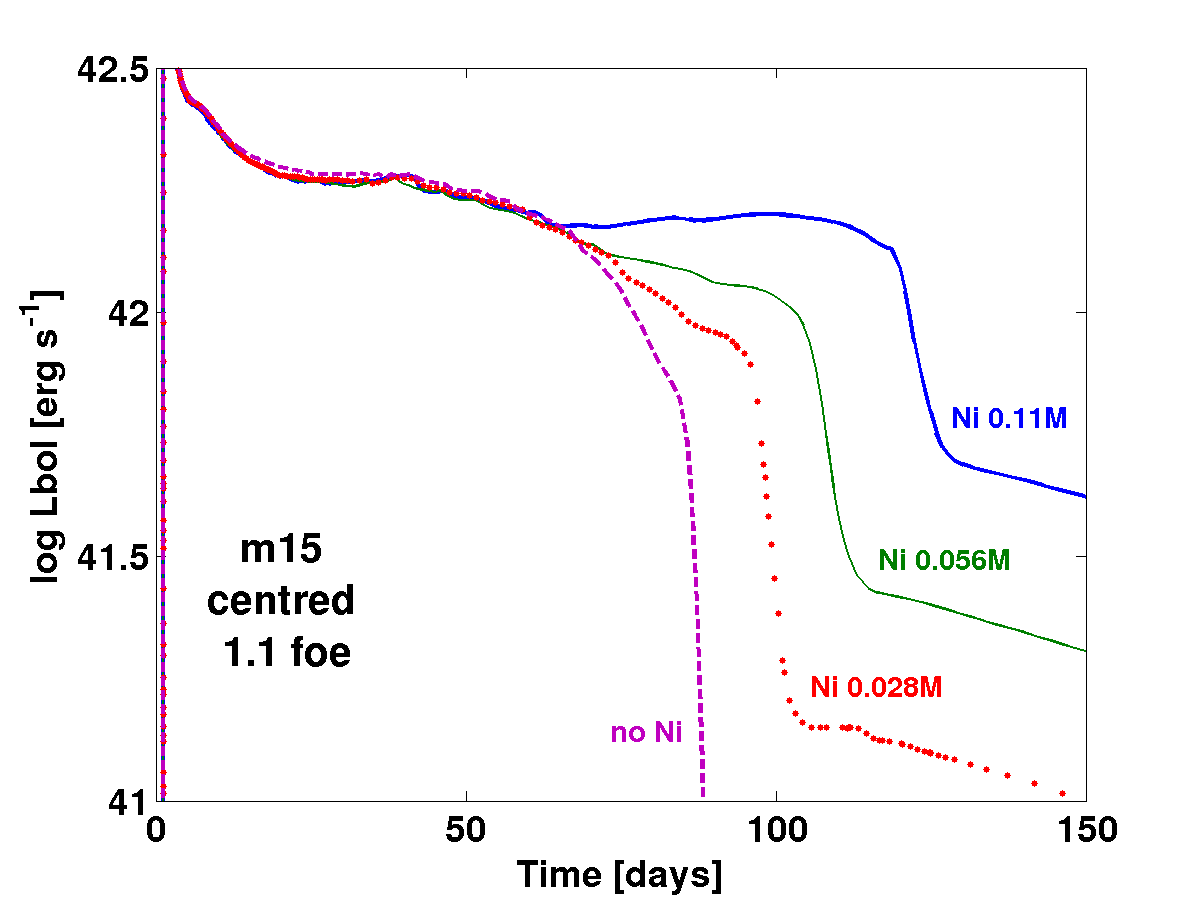}~
\includegraphics[width=0.5\textwidth]{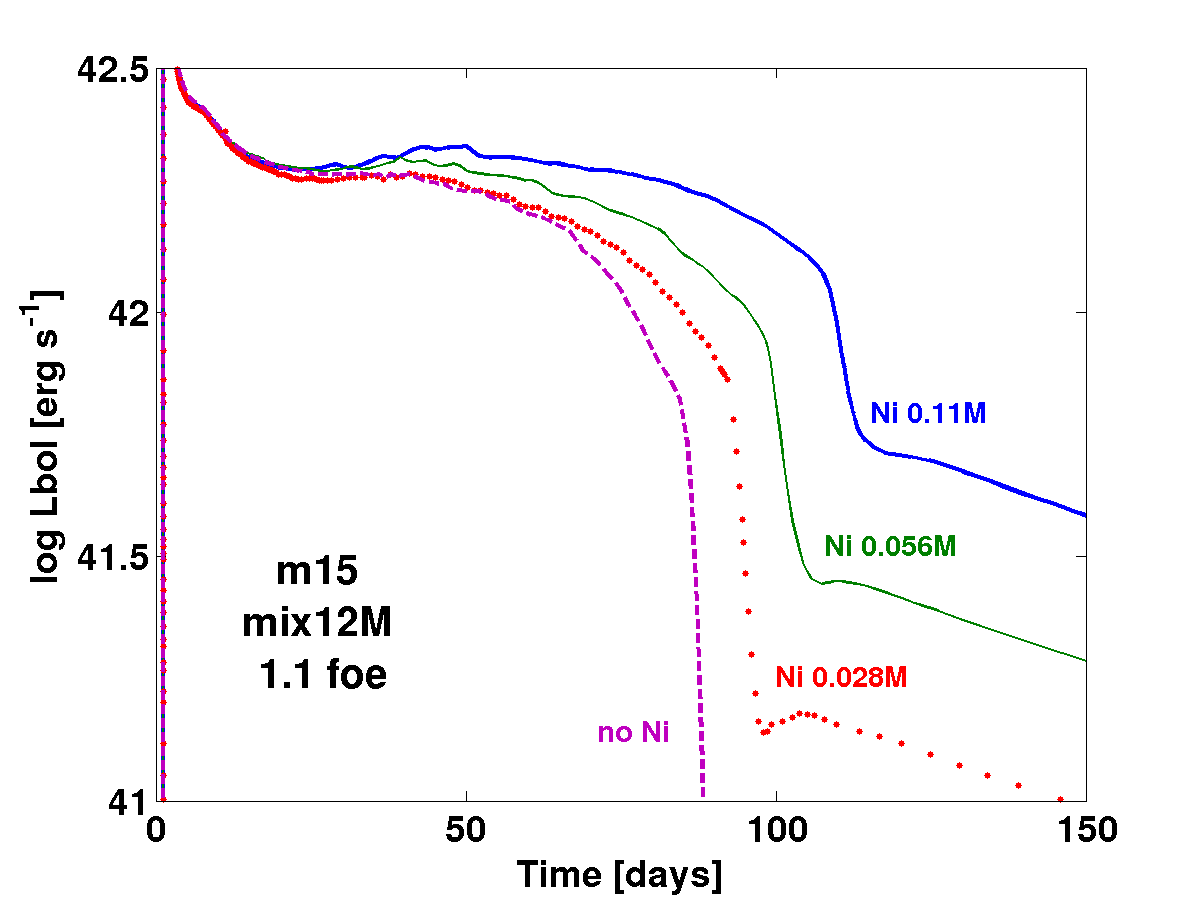}
\caption{Bolometric light curves (same as in figures \ref{figure:m12lbol} and \ref{figure:m15lbol}) where in each panel the amont of \Ni{} varies while the mixing and the energy reamins constant.}
\label{figure:diffNi}
\end{figure*}

In Figures~\ref{figure:m12lbol}, \ref{figure:m15lbol} and \ref{figure:diffNi}, we present the resulting bolometric light curves for the model m12
and the model m15 with different amount of \,\,$^{56}$Ni mass, degree of nickel mixing, and explosion
energy. The labels along the curves indicate the explosion energy and
corresponding average parameter $\eta_\mathrm{Ni}$.  From these figures we can see several clear features. 

First, the time at which \Ni{} starts to affect the light curve is determined
only by the degree of mixing and is independent of the \Ni{} mass (see figure
\ref{figure:diffNi}).  A centrally concentrated \Ni{} starts affecting the
emission around the time the plateau ends in the no \Ni{} light curve
while a fully mixed \Ni{} has an affect already from the beginning of the
plateau.  In all cases, once \Ni{}  starts affecting it increases the
luminosity leading, as expected, to a flatter and longer plateau.  The
prominence of the \Ni{} emission is increased with the \Ni{} mass and reduced
with the explosion energy.  In general, for a given type of mixing, light
curve with similar values of $\eta_\mathrm{Ni}$ (although different amounts
of \Ni{} mass and explosion energy) show similar effect of \Ni{} on the light
curve.  The effect is of course more prominent for higher $\eta_\mathrm{Ni}$ values.

An interesting property of central \Ni{} mixing is that the light curve can be
roughly separated between the cooling emission and the \Ni{} driven radiation. 
The luminosity of each phase depends on different properties of the
progenitor, therefore, they are not necessarily similar.  Indeed, in some
of the light curves the transition between the two phases can be seen.  The
nature of this transition depends on $\eta_\mathrm{Ni}$.  For very low
values of $\eta_\mathrm{Ni} \lesssim 0.1$ the effect of \Ni{} can be hardly
identified.  For slightly higher values, but still relatively low, $\eta_\mathrm{Ni}
\sim 0.2$, the \Ni{} contribution becomes apparent, but is still less luminous
then earlier cooling emission, therefore, near the middle of the plateau, i.e. around day~50,
the decay of the light curve becomes steeper.  For intermediate values
$\eta_\mathrm{Ni} \simeq 0.5$ the \Ni{} phase has comparable contribution as the cooling emission, and
the transition between the two phases can be hardly observed.  Finally, for
$\eta_\mathrm{Ni} \gtrsim 1$ the \Ni{} driven emission is seen as a clear
``bump'' that starts rising from the middle of the plateau.  Such a bump was
never seen in a type II SN from a red supergiant.  In simulations where \Ni{}
is mixed out to the envelope its contribution is also mixed with that of the
cooling emission leading to a smoother plateau evolution with no observable
transition for any value of $\eta_\mathrm{Ni}$.  A result of the difference
between the mixing levels is  that a more concentrated \Ni{} leads to
longer plateaus than less concentrated \Ni{}, while the flattening is more
prominent when \Ni{} is mixed to outer layers.  Light curves for the
``boxcar'' mixed models look similar to the models with
the centrally located $^{56}$Ni (for a given $^{56}$Ni mass) during the
plateau phase. This happens because
the major fraction of $^{56}$Ni in ``boxcar'' mixed models is located in the centre
(see Section~\ref{subsect:mix} and Table~\ref{table:models}), and 
a little mass of $^{56}$Ni is contained in the ejecta at higher velocity. 

Interestingly, the values of $\eta_\mathrm{Ni}$ in most of the sample
explored by \citet{2016ApJ...823..127N} are  in the range 0.3\,--\,0.7, which
implies a non-negligible \Ni{} contribution.  However, for these values it is
hard to determine the type of \Ni{} mixing based on the light curve alone. 
The reason is that this is exactly the values were the \Ni{} phase in
the models with centrally concentrated \Ni{} continues the cooling emission phase
smoothly with no obvious observational feature.  The only SN for which our
results can strongly constrain the mixing is SN\,2009ib for which
$\eta_\mathrm{Ni}=2.6$.  The light curve of SN\,2009ib shows a long 130-day very smooth
plateau \citep{2015MNRAS.450.3137T}.  This light curve is very different than those that we see in our
simulations where the \Ni{} is concentrated in the centre.  In fact, it seems
that for $\eta_\mathrm{Ni}=2.6$ even in cases where \Ni{} is only partially
mixed into the envelope we should identify the time at which the \Ni{}
contribution kicks in.  Among our light curves with similar
$\eta_\mathrm{Ni}$ values only those with \Ni{} mixing throughout the envelope
resemble the light curve of SN\,2009ib. 

We notice that bolometric light curves for some models
have a specific step-like feature during transition from
the standard recombination phase to the radioactive tail. 
These are models with either low explosion energy combined with low mass of $^{56}$Ni
located in the centre (see Figure~\ref{figure:m12lbol}, the model m12, 0.025~\Msun of $^{56}$Ni, 0.4~foe, blue solid curve)
or models with ``boxcar'' mixing and low or medium explosion energy (see
Figure~\ref{figure:m15lbol}, the model m15, e.g. 0.028~\Msun{}, 0.53~foe and
1.1~foe, green and blue dashed curves). We noticed, that the step disappears
for the mentioned models at higher
explosion energy (e.g. with 1.53~foe in the model m15).
In those models, the step is caused by helium recombination, i.e. when
photosphere recedes to the helium shell. Photons produced in $^{56}$Ni and
$^{56}$Co decay diffuse and ionise helium, therefore, keeping photosphere at
this layer for a while. The step does not appear in cases of relatively
higher energy, because the overall internal energy is higher and relative
contribution from Ni-heating is lower. Similarly, the transition from
hydrogen to helium recombination is smooth in the models with stronger Ni-mixing.
In fact, in these cases Ni-produced photons heat hydrogen-rich atmosphere
equally as helium layer (in uniform mixing). While hydrogen ionisation
supports photosphere at larger radius, the whole ejecta expands and then
photosphere quickly drops through helium layer which becomes transparent.
Nevertheless, the majority SNe~IIP have
no such shape in their light curves. Thus, our results suggest that
all mentioned models which are attributed by the step-feature in the
bolometric light curve are not the progenitors for normal SNe~IIP. This result however is still needed to be confirmed by other numerical codes to verify that the observed step-like feature is indeed seen in these models. If confirmed then ``boxcar'' mixed models with low (about 0.5~foe) and
intermediate (1~foe) explosion energy could not reproduce normal type~IIP
supernova light curves, while uniformly mixed models do.

As our study is focused on the light curve, we present
ejecta properties on the coasting phase in Appendix~\ref{appendix:append}. Namely, we chose the
model m12 with 0.045~\Msun{} of \Ni{} and
the model m15 with 0.056~\Msun{} of \Ni{}, and show selected
species, hydrogen, helium, oxygen, silicon, and iron, presented in the SN
ejecta along velocity.

\subsection[Extension of the plateau]{Extension of the plateau duration}
\label{subsect:duration}

Our goal here is to quantify the effect of \Ni{} on the plateau duration $T_{pl}$, finding its dependence of the observable $\eta_\mathrm{Ni}$ 
and on the \Ni{} mixing. 
We measure $T_{pl}$ as time since explosion till the middle of transition between plateau and the radioactive tail. According to \citep{1991SvAL...17..210C}, duration of plateau corresponds to time when the recombination front traverses throught the SN ejecta.
The plateau extension was previously studied. \citet{2009ApJ...703.2205K} and \citet{2016ApJ...821...38S}
derived the following relation between the duration of the
plateau, the explosion energy and the progenitor properties based on a set of numerical runs (Eq.~19 of \citet{2016ApJ...821...38S}):

\begin{equation}
\frac{T_{pl}}{T_{pl}(Ni=0)} = \left( 1 + C_f\,
\frac{M_\mathrm{Ni56}}{E_{51}^{1/2} M_{ej,10}^{1/2} R_{0,500}} \right)^{1/6} \,.
\label{equation:Kasen}
\end{equation}

where $C_f$ is a constant that depends on the progenitor structure, $E_{51}$
is the explosion energy in units of foe, $M_{ej,10}$ is the ejecta mass is
units of 10~\Msun{} and $R_{0,500}$ is the progenitor radius in units of
500~\Rsun{}.  Interestingly, for a given progenitor structure $\eta_\mathrm{Ni}
\propto \frac{M_\mathrm{Ni56}}{E^{1/2} M_{ej}^{1/2} R_{0}}$  (see
Section~\ref{sect:eta}), therefore, we fit the results of our simulations to the
relation

\begin{equation}
\frac{T_{pl}}{T_{pl}(Ni=0)} = ( 1 + a \,\eta_\mathrm{Ni})^{\,1/6} \,.
\label{equation:KasenEta}
\end{equation}

\begin{figure*}
\centering
\includegraphics[width=.9\textwidth]{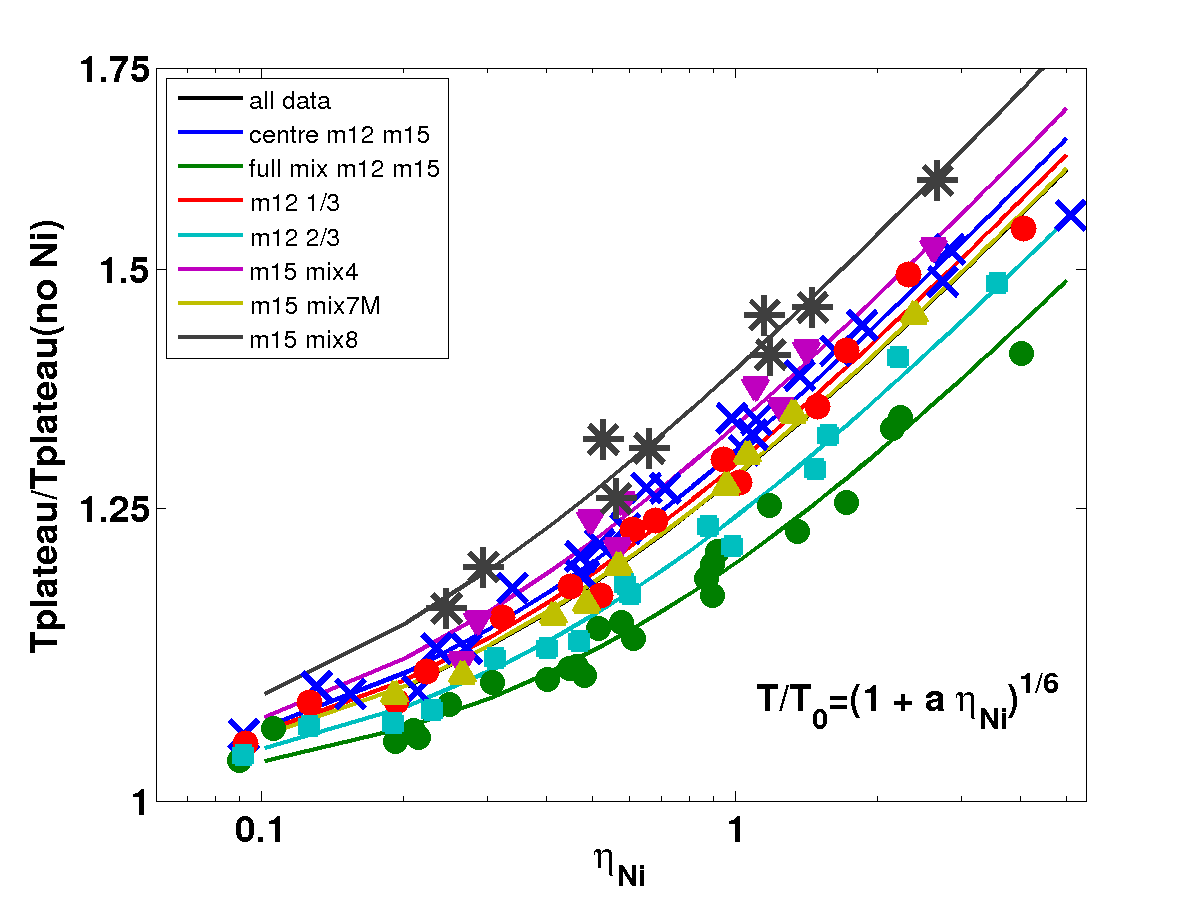}
\caption{Plateau duration for models with $^{56}$Ni relative to the models without $^{56}$Ni
along parameter $\eta_\mathrm{Ni}$.}
\label{figure:TTpl_eta}
\end{figure*}

The results are shown in Figure~\ref{figure:TTpl_eta}.  We use different
values of $a$ for different mixing types, where the range of $a$ values that
we find is between 2 and 6.
Firstly, it is clear that formula~\ref{equation:KasenEta} provides
an excellent fit to the data.  Its main advantage over equatio~\ref{equation:Kasen} is that $\eta_\mathrm{Ni}$ is an observable,
therefore, it can be measured for any SNe with a good bolometric light curve. 
Secondly, it shows the dependence of the plateau extension on the mixing. 
Maximal extension is obtained for the boxcar ``mix8'' and ``mix4'' mixings
followed closely by the centrally concentrated \Ni{}.  For these types of
mixing $a \approx 5-6$.  The plateau extension drops when \Ni{} is heavily
mixed into the envelope, and the smallest effect is measured for fully
mixed \Ni{} where $a \approx 2$.  Finally, for our two progenitors
we found that similar mixing types resulted in the same value of $a$. 
Although we examined only two progenitor models this suggests that most of
the dependence of $C_f$ on the progenitor structure in  equation
\ref{equation:Kasen} is absorbed into the parameter $\eta_\mathrm{Ni}$ and that
the coefficient $a$ depends mostly on the mixing type.
Hence, we suggest the averaged formula which
is valid with roughly 10\,\% accuracy:
\begin{equation}
\frac{T_{pl}}{T_{pl}(Ni=0)} = ( 1 + 4 \,\eta_\mathrm{Ni})^{\,1/6} \,.
\label{equation:KasenEtaAnumber}
\end{equation}

Applying these results to the sample of \citet{2016ApJ...823..127N}  we find
that for typical explosions with $\eta_\mathrm{Ni} \approx 0.5$ the \Ni{} extends
the plateau by 15\,\%\,--\,25\,\%, depending on the type of mixing.  As the typical
observed plateau duration is about 100~days, this implies that if there was
no \Ni{} the typical plateau duration would have been about 80~days.  In the
case of SN\,2009ib where the plateau is unusually long, we find that for
$\eta_\mathrm{Ni}=2.6$ and a uniform mixing throughout the envelope (as
inferred from the light curve shape) the \Ni{} extends the plateau by 35\,\%. 
Thus, the unusual plateau length of SN\,2009ib is mostly due to \Ni{}, as without \Ni{} the
plateau duration would have been shorter than 100~days.

\subsection{Mitigating the plateau decline rate}

\begin{figure*}
\centering
\includegraphics[width=.33\textwidth]{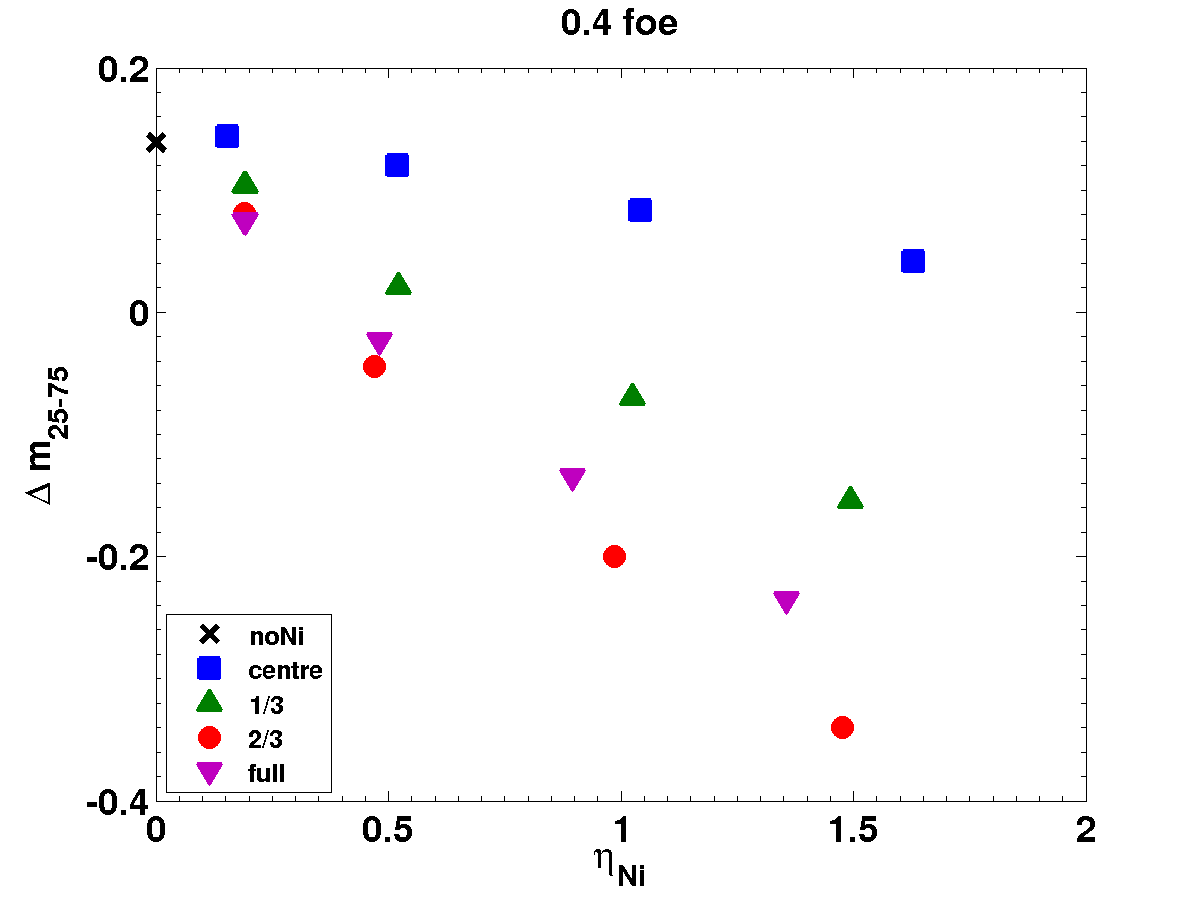}~
\includegraphics[width=.33\textwidth]{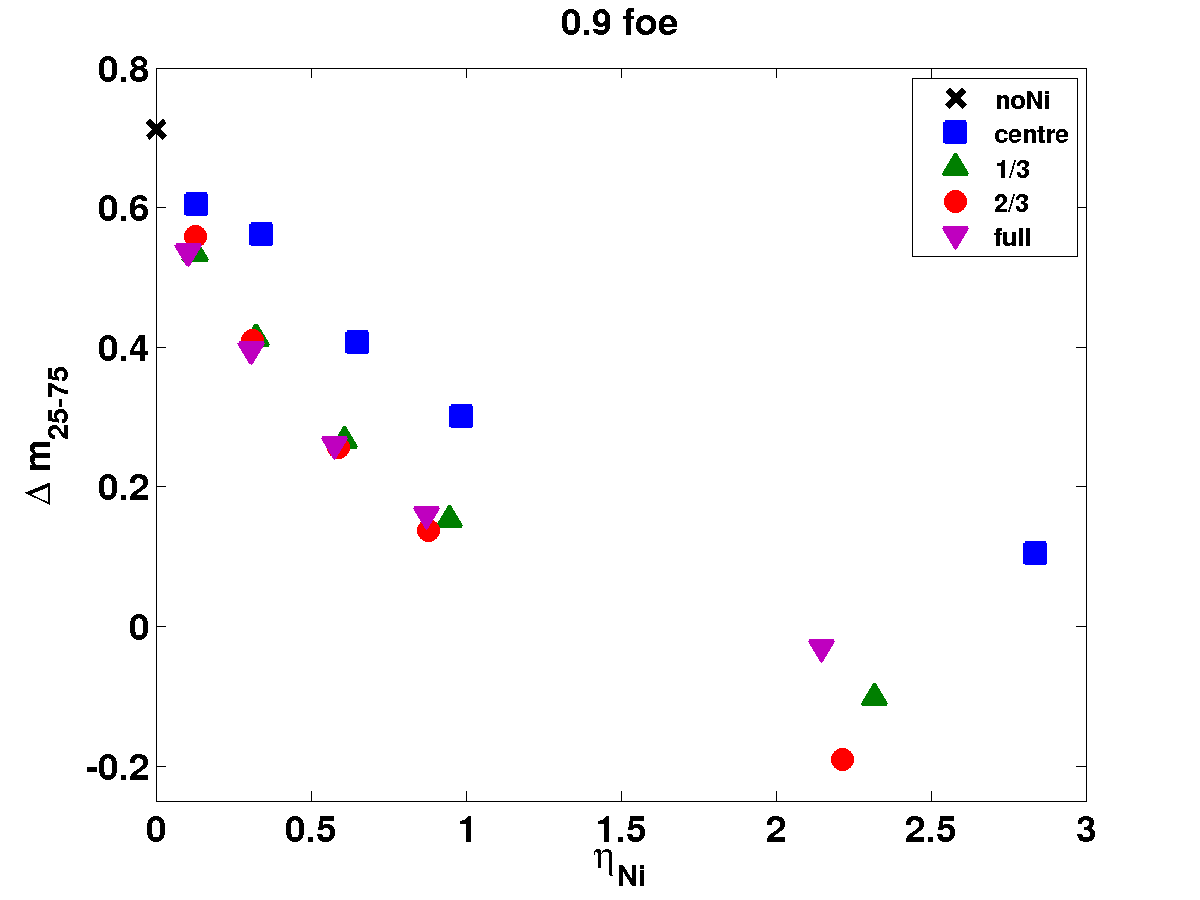}~
\includegraphics[width=.33\textwidth]{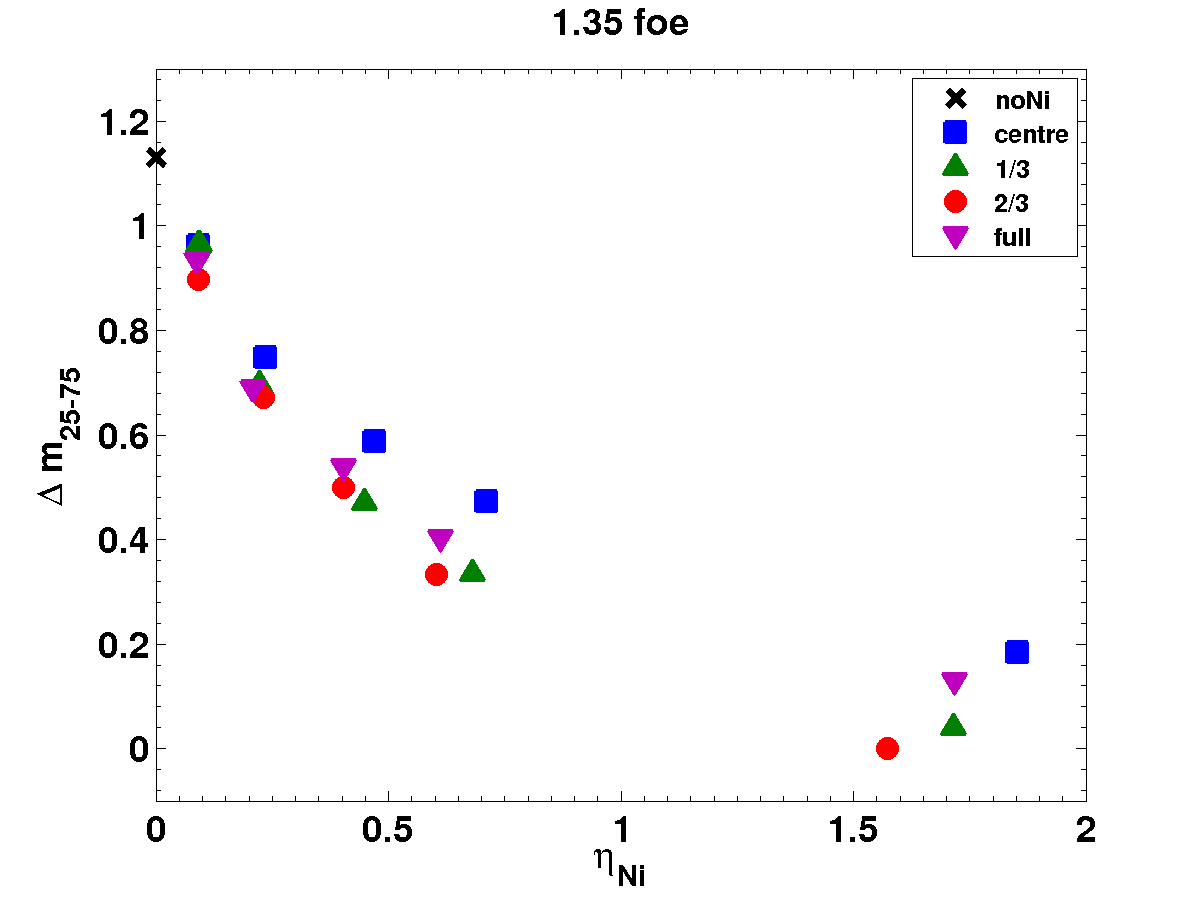}
\caption{The effect of \Ni{} on the decline rate of explosions with different energies for diffrent mixing types.} 
\label{figure:etaNi_dm}
\end{figure*}

Another effect of \Ni{}, when it is present in the SN ejecta, is 
increasing the plateau luminosity
at late time, thereby reducing the decline rate, i.e., making the light curve flatter. 
We use the bolometric increase in magnitude (i.e., drop in luminosity)
between days 25 and 75, $\Delta m_{25-75}$, as a measure of the decline
rate.  The values in the \citet{2016ApJ...823..127N} sample vary between
$\Delta m_{25-75}=-0.1$ for light curves that show slow brightening during
the plateau to $\Delta m_{25-75}=1.2$ (i.e., a decline rate of 2.4 mag/100
day) for fast declining SN, which are usually classified as type IIL. 
\citet{2016ApJ...823..127N} test for a correlation between
$\eta_\mathrm{Ni}$ and $\Delta m_{25-75}$ finding a significant
anti-correlation, suggesting that \Ni{} is responsible for at least some of
the plateau flatness.  They continue with an attempt to measure this effect,
estimating that in SNe with flat plateaus \Ni{} contributes about 1 mag/100
day to the plateau (i.e., without \Ni{} these SNe would have showen a decline
rate of $\Delta m_{25-75} \approx 0.5$).

Figure \ref{figure:etaNi_dm} shows $\Delta m_{25-75}$ as a function of
$\eta_\mathrm{Ni}$ for progenitor m12 (the results for m15 are very
similar).  The different panels are for different explosion energies. 
First, we find that without \Ni{} more energetic and luminous explosions
evolve faster and have a faster decline rate.  This is consistent with the
correlation found between SN luminosity and the decline rate
\citep{2014ApJ...786...67A,2014MNRAS.445..554F}, suggesting that high
explosion energy might be at least one of the reasons for the fast decline
observed in some SNe\,IIL.  The effect of \Ni{} on the decline rate is also seen
clearly.  For all mixing types higher \Ni{} mass results in slower
decline rates.  This effect is seen for all mixing types, although it is
less prominent when \Ni{} is concentrated in the center.  Quantitatively, for
$\eta_\mathrm{Ni} \approx 0.5$  the effect on low energy explosion is minor,
but on explosions with energy of about 1~foe it reduces the decline rate by
about 1~mag/100~days (i.e., reducing $\Delta m_{25-75}$ by 0.5).  For
$\eta_\mathrm{Ni}=0.1$, the effect is minor, and the resulting light curve
decay is similar to the light curve without \Ni{}. For $\eta_\mathrm{Ni}>1$, the plateau is
always very flat, which is consistent with the light curve of SN\,2009ib,
and in some cases even slowly rising.


\subsection{Correlations between $\eta_\mathrm{Ni}$ and the drop from the plateau to the radioactive tail}
\label{subsect:correlations}

\begin{figure}
\centering
\includegraphics[width=0.5\textwidth]{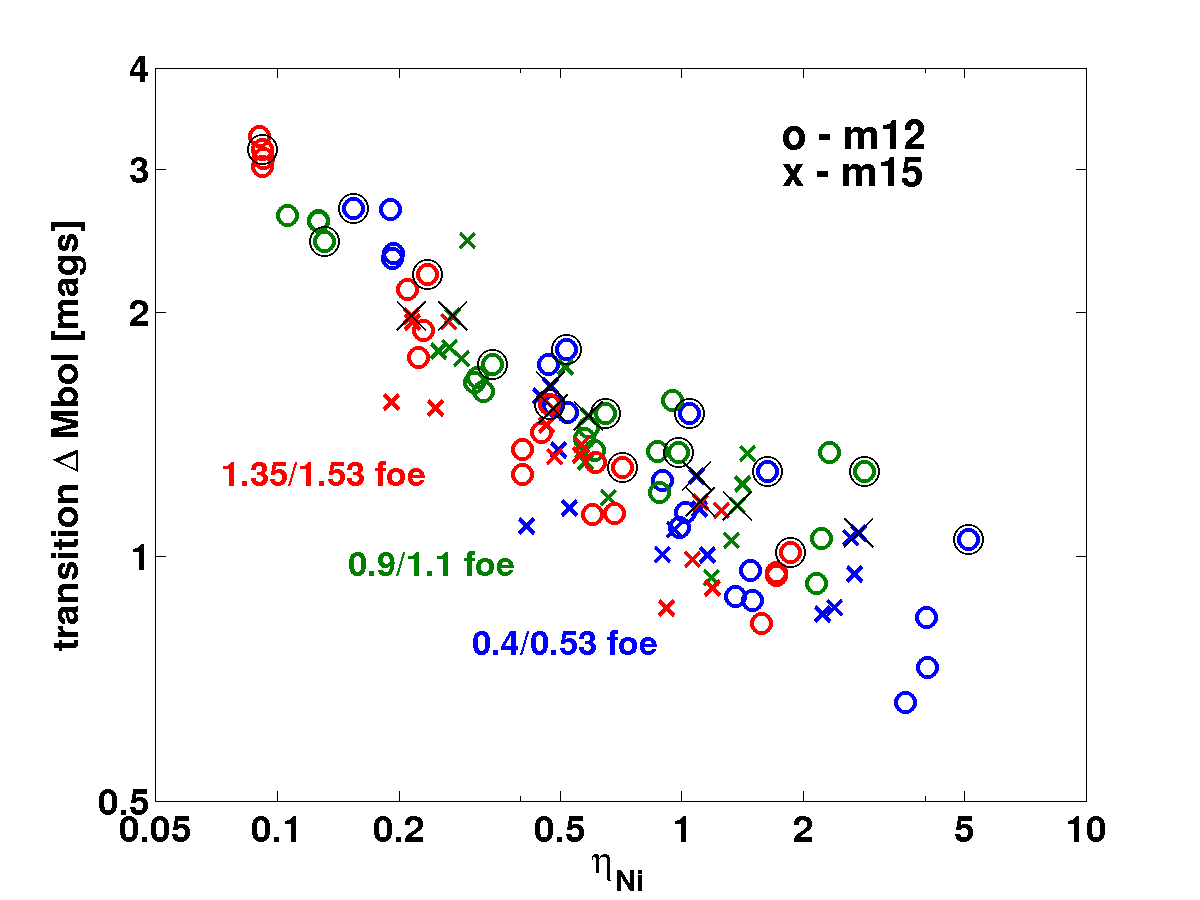}~
\caption{Correlation between drop of the bolometric light curve
during transition from plateau and tail and $\eta_\mathrm{Ni}$. We marked central cases with additional black light
circles and crosses for the models m12 and m15, respectively.}
\label{figure:correlation}
\end{figure}

Figure~\ref{figure:correlation} shows the drop in the bolometric light
curve during the transition from the plateau to the tail
as a function of $\eta_\mathrm{Ni}$. 
We define the transition as a difference in bolometric magnitude between
the end of plateau (when the plateau slope starts changing noticeably) and the beginning of radioactive tail.
It shows a rather tight correlation with $\eta_\mathrm{Ni}$ while the scatter is mostly due to the different mixing types.
The transition between the plateau and the
tail is a complicated characteristic which depends on the explosion energy,
global progenitor properties and the mass of
$^{56}$Ni and its distribution, i.e. $\eta_\mathrm{Ni}$. 

\subsection[Photospheric velocity]{Photospheric velocity}
\label{subsect:uph}

\begin{figure}
\centering
\includegraphics[width=0.5\textwidth]{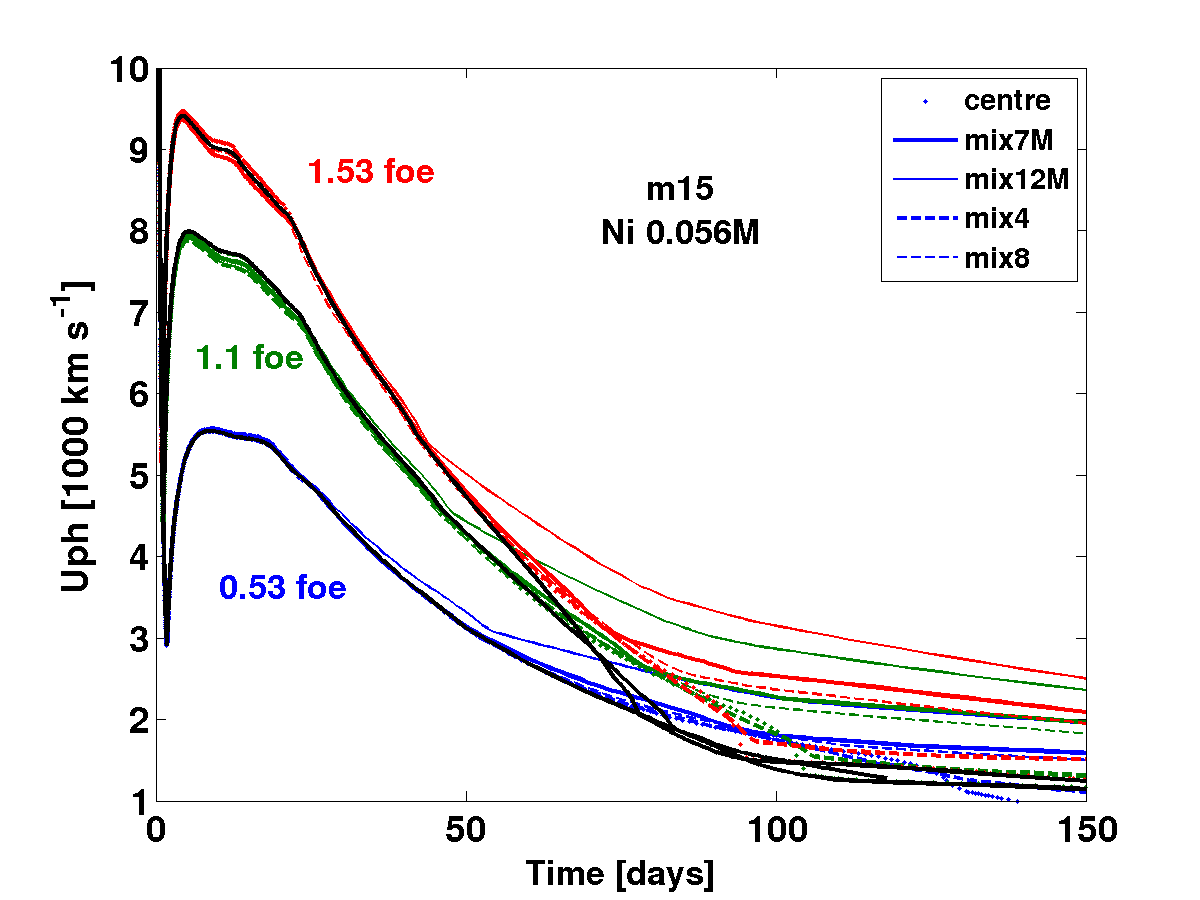}
\caption{Photospheric velocity evolution for the model m15 with
0.065~\Msun{} of $^{56}$Ni and different explosion energies, 0.53~foe (blue),
1.1~foe (green), and 1.53~foe (red). Black curves represent photospheric
velocity evolution for the model m15 with no \Ni{} included.}
\label{figure:uphm15ni05}
\end{figure}

Theoretically, we expect that $^{56}$Ni presented in the
supernova ejecta heats and ionises matter while supplying high energy
photons and positrons, therefore, keeping the
photosphere at larger radii where velocity is higher.
We define photospheric velocity as velocity of the Lagrangian zone where the integrated Rosseland optical depth is equal 2/3.
In Figure~\ref{figure:uphm15ni05}, we depicts the evolution of the photospheric velocity
models m15 with 0.065~\Msun{} of $^{56}$Ni and different explosion energies, 0.53~foe,
1.1~foe, and 1.53~foe. It shows that the main factor that determines the photospheric 
velocity is the explosion energy. 
Since the typical ejecta velocity scales as $\sqrt{E/M}$, we expect the ejecta mass to have a 
similar effect as the explosion energy \citep[e.g.,][]{1993ApJ...414..712P}.
The effect of \Ni{} on the photospheric velocity is significant only at late times. Similarly to the light 
curve, higher level of mixing starts affecting the velocity at earlier time.
However, this happens later than in the light curves. At day~50, only full 
mixing shows some \Ni{} effects of order 10\,\%. Towards the end of the plateau 
all mixing types affects the velocity, where in the case of uniform mixing the 
line velocity can be more than twice the velocity at the same time without \Ni{}. 
To conclude, $^{56}$Ni has a negligible contribution to the photospheric
velocity evolution up to the middle of the plateau, and starts playing some role by the end
of plateau. However, it is expected that even small changes in photospheric velocity due to Ni-heating
will be seen in spectra.

\begin{figure}
\centering
\includegraphics[width=0.5\textwidth]{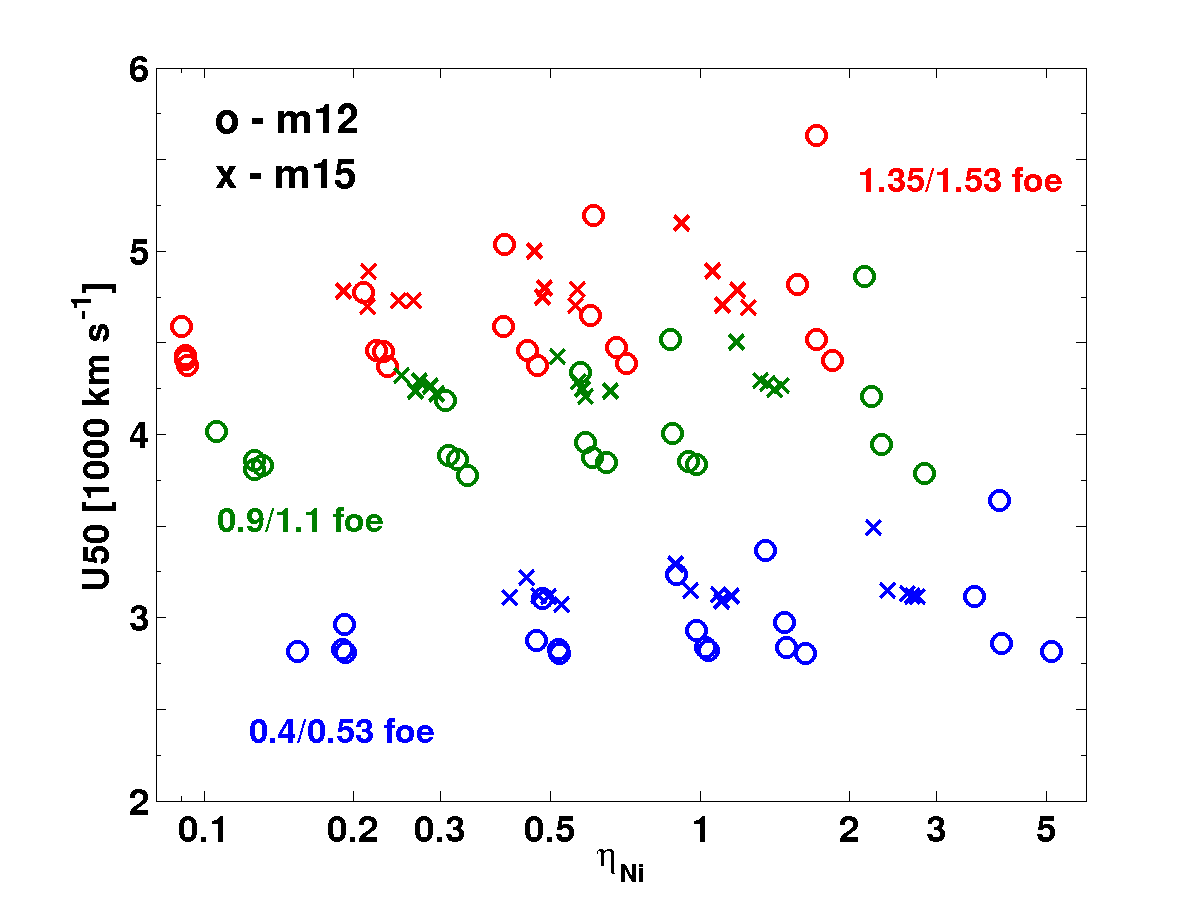}
\caption{Photospheric velocity at day~50 for all models in the study.}
\label{figure:uph50}
\end{figure}

Since the velocity at day~50 is often used to characterise the explosion
energy and ejecta mass, we present the photospheric velocity at day~50 in
Figure~\ref{figure:uph50}, for all models in the study along the parameter
$\eta_\mathrm{Ni}$.  We show that there is no major effect, and for
$\eta_\mathrm{Ni} \approx 0.5$ the photospheric velocity is at most 10\,\%
faster.

\subsection[Colour \textit{B-V}]{Colour \textit{B-V}}
\label{subsect:colour}

\begin{figure}
\centering
\includegraphics[width=0.5\textwidth]{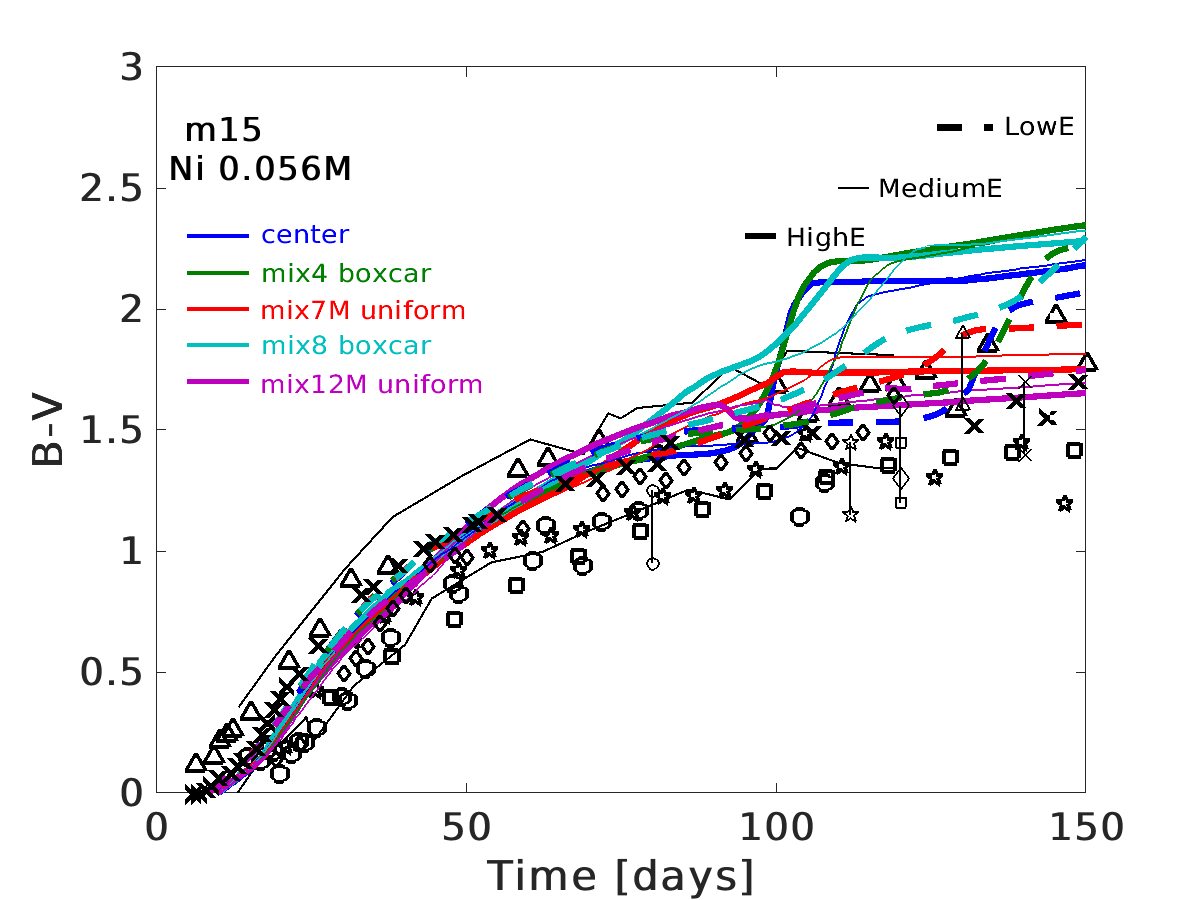}
\caption{\emph{B}--\emph{V} colour for the
model m15 with 0.056~\Msun{} of \,\,$^{56}$Ni mixed differently and
different explosion energy, 0.53~foe (``LowE'', thick solid), 1.1~foe
(``MediumE'', thin solid), and 1.53~foe (``HighE'', thick dashed). Superposed 
symbols are for SNe\,IIP from 
\citet{2014MNRAS.445..554F,2014MNRAS.442..844F} and \citet{2016MNRAS.459.3939V}.
Among others, crosses stand for 1999em, triangles stand for 1999gi, stars
stand for 2001X, diamonds stand for 2005ay, circles stand for 2013fs, and squares
stand for ASASSN14ha. 
Vertical lines indicate the end of plateau phase in each individual 
SNe with symbols corresponding to the colour data.
Vertical bars indicate the end of plateau for given SN. Symbols at the end
of a bar correspond to the marker type of the curve.}
\label{figure:color}
\end{figure}

In Figure~\ref{figure:color}, we plot \textit{B}--\textit{V} colour
for model m15 with 0.056~\Msun{} of \,\,$^{56}$Ni with various mixing types and
explosion energies of 0.53~foe (``LowE'', thick solid), 1.1~foe
(``MediumE'', thin solid), and 1.53~foe (``HighE'', thick dashed). We compare our theoretical curves with a few
normal SNe~IIP: 1999em (crosses), 1999gi (triangles), 
2001X (stars), 2005ay (diamonds), 2013fs (circles), ASASSN14ha (squares) \citep{2014MNRAS.445..554F,2016MNRAS.459.3939V}, 
and averaged \textit{B}--\textit{V}-curves \citep[thin black curves,][]{2014MNRAS.445..554F}.
We find that models with centrally concentrated $^{56}$Ni (blue curves) and ``boxcar'' mixing
(green and cyan curves) quickly change their \textit{B}--\textit{V} colour
during the transition from the plateau to the radioactive tail.  This is in
contrast to the observations that demonstrate monotonic evolution of
\textit{B}--\textit{V} colour.  All the models with uniformly mixed
$^{56}$Ni show evolution that is similar to the observed one.

Our results suggest that the \Ni{} is well mixed into the envelope 
in normal SNe~IIP.
Note however, that \verb|STELLA| produces reliable
evolution of radiation field for supernova ejecta when a large fraction of
the ejecta is optically thick, i.e. during photospheric phase. 
By the end of transition from the plateau to the tail, the ejecta become optically thin and
the photosphere recedes deep into the ejecta. At this time
the overall bolometric light curves predicted by \verb|STELLA| are also reliable.
However, lines begin 
playing a significant role at late time, and \verb|STELLA| colours are less reliable. 
Therefore, detailed simulations with non-thermal effects
and larger atomic data base (like it is done in \verb|CMFGEN|, \citet{2012MNRAS.426.1671L})
are needed to confirm our result about the \textit{B}--\textit{V} colour evolution.
Previously, \citet{2011MNRAS.410.1739D} present non-LTE simulations
for two models. Their \textit{B}--\textit{V}-colour curves lie below the
observed range marked in our Figure~\ref{figure:color}. Nevertheless, their
light curves are not properly reproduce normal SNe~IIP. \citet{2009ApJ...703.2205K}
present broad band light curves for a particular model with applied boxcar
mixing. The derived \textit{B}--\textit{V} colour is in very good agreement
with our results for the ``mix4'' and ``mix8'' boxcar curves, i.e. the colour
reddens in a step-like way during the transition to the radioactive tail.
Note though that \citet{2009ApJ...703.2205K} do not include non-LTE effects
in their study.

\subsection[The influence of different opacity treatment]{Comparison to V1D}
\label{subsect:compare}

\begin{figure}
\centering
\includegraphics[width=0.5\textwidth]{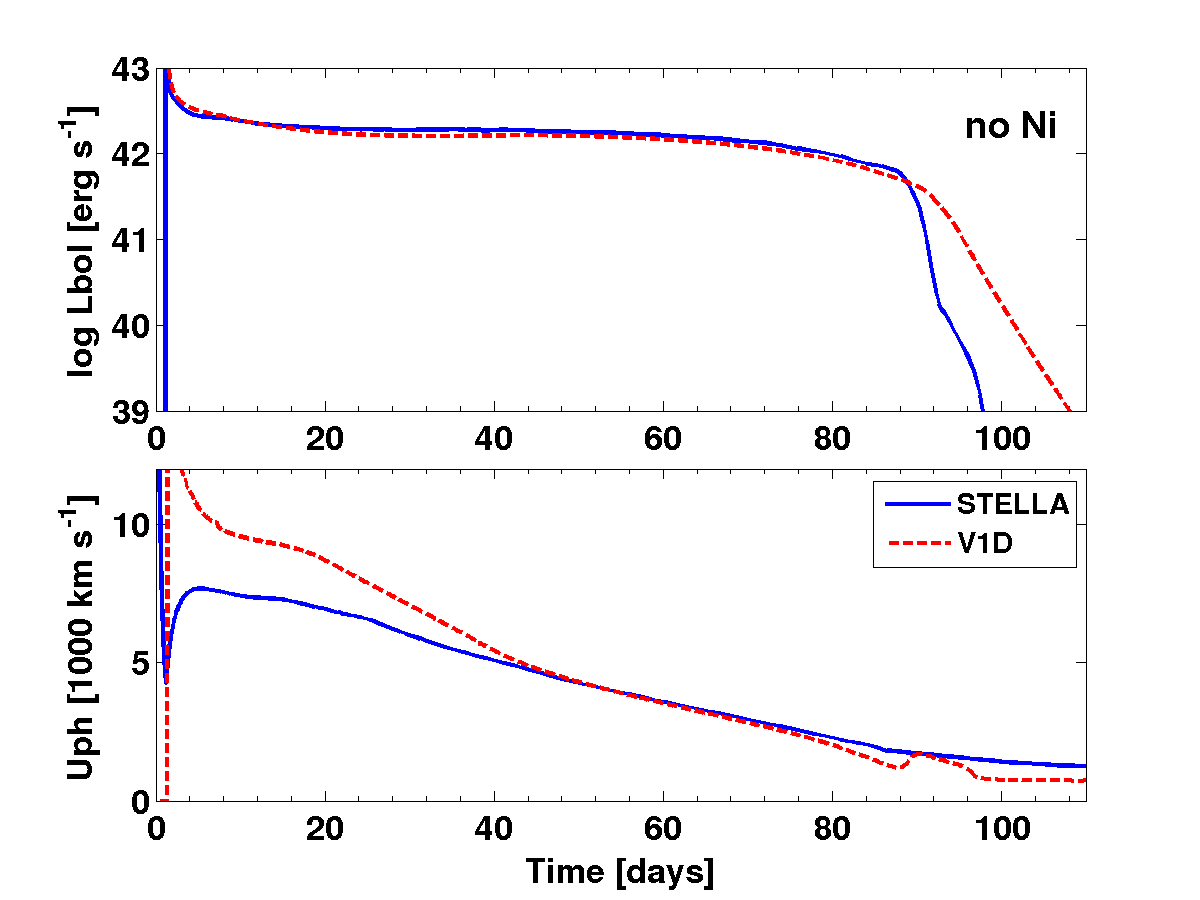}\\
\includegraphics[width=0.5\textwidth]{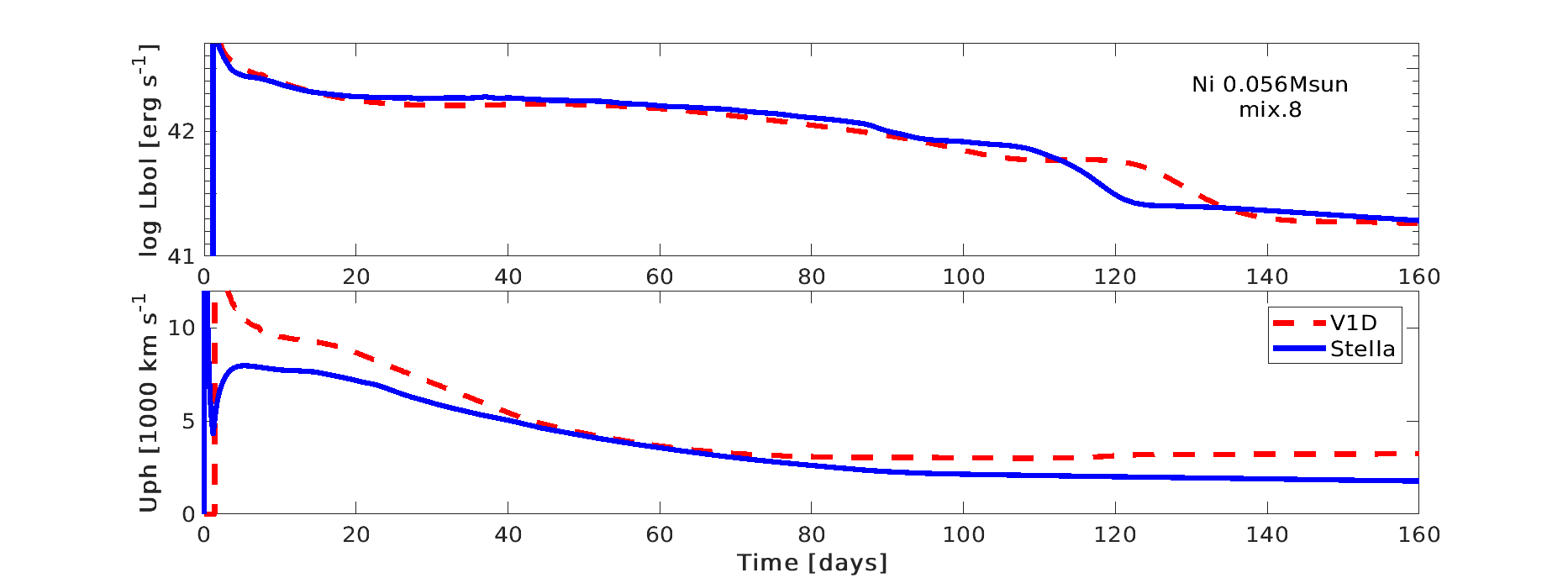}
\caption{V1D (red dashed) and STELLA (blue solid) bolometric light curves and 
photospheric velocity for the model m15 with no radioactive material included
(top) and with 0.056~\Msun{} if \,\,$^{56}$Ni mixed in 80\% of ejecta (bottom).}
\label{figure:compare}
\end{figure}

The opacity treatment remains the core aspect in the radiative transfer
simulations which provide an uncertainty in the resulting data.
Figure~\ref{figure:compare} demonstrates that using different sets of
lines and different assumptions about thermodynamical equilibrium (LTE
versus non-LTE) leads to visible differences in bolometric light curves.

In this section we compare simulations done with \verb|STELLA| and \verb|V1D|.
\verb|V1D| uses opacity tables compiled from the \verb|CMFGEN| data 
\citep{2010MNRAS.405.2141D,2010MNRAS.405.2113D,2015MNRAS.449.4304D} which is
significantly broader than the \verb|STELLA| standard settings. \verb|CMFGEN|
includes about 500,000 lines and level populations without assumption of
LTE, and treats non-thermal excitation, while \verb|STELLA| has 160,000 lines and computed level populations
based on modified Saha equations. From the comparison plots in
Figure~\ref{figure:compare}, it is obvious that the effective \verb|V1D|
opacity is larger than the opacity in \verb|STELLA|. This makes the \verb|V1D| plateau
7-day longer even in the model without radioactive nickel \,\,$^{56}$Ni
(upper panel of Figure~\ref{figure:compare}).
Apart from the plateau duration, luminosity on the plateau varies. Hence, \verb|V1D|
predicts a 0.04~dex dimmer plateau compared to \verb|STELLA| in the case of
0.056~\Msun{} of \,\,$^{56}$Ni (bottom panel of Figure~\ref{figure:compare}).
Nevertheless, the qualitative agreement between \verb|V1D| and \verb|STELLA|
is very good.

Photospheric velocity estimated by \verb|V1D| and \verb|STELLA| is
different, as seen in Figure~\ref{figure:compare}, while
photospheric temperature and radius are in very good agreement. \verb|V1D|
overestimates the velocity by 2000\,km\,s$^{\,-1}$ during the post-SBO cooling
phase and up to the middle of plateau, and by 1000\,km\,s$^{\,-1}$ after the
middle point. We explain this difference by the way \verb|V1D| solves the radiative
transfer equations, i.e. diffusion approximation.

%% file: conclusion3.tex

\section[Conclusions]{Conclusions} 
\label{sect:conclusions}

In the present study, we carried out a systematic analysis of the impact of
$^{56}$Ni presented in the SN ejecta on the resulting light curves of SNe\,IIP.
For this, we used two red supergiant models computed with \verb|MESA|. We
vary amount of \Ni{} (0.01\,--0.14~\Msun{}) and its mixing (centrally
concentrated to full), and explosion energy
(0.4~foe to 1.53~foe) and computed a set of light curves with \verb|STELLA|. 
Our light curves are available via link \url{https://wwwmpa.mpa-garching.mpg.de/
ccsnarchive/data/Kozyreva2018/}.

Based on our light curve simulations, we conclude that even small amount of radioactive nickel $^{56}$Ni
presented in the supernova ejecta noticeably modifies the bolometric light curve.
There is a combination of two effects from Ni-heating on the plateau duration and
shape: (1) radioactive nickel \Ni{} extends the plateau, and (2) $^{56}$Ni
flattens plateau decline rate.

To evaluate the importance of \Ni{} impact, we used the parameter
$\eta_\mathrm{Ni}$ which is a ratio between time weighted \Ni{} deposited energy
and the weighted shock deposited energy. We found that the extension of the plateau due to presence
of \Ni{} can be accurately approximated by a simple formula with an argument $\eta_\mathrm{Ni}$
(Equation~\ref{equation:KasenEta} and Figure~\ref{figure:TTpl_eta}).
We found that in most observed type IIP SNe the plateau is extended by 15-25\,\%.
We also found that \Ni{} effectively flattens the plateau decline. In fact, the drop
in light curves between day~25 and day~75 is correlated with
$\eta_\mathrm{Ni}$, i.e. decline rate is lower (a light curve is flatter) for
higher $\eta_\mathrm{Ni}$. For an intermediate explosion
energy of about 1~foe and $\eta_\mathrm{Ni} \approx 0.5$ the decline rate is reduced by about 1~mags/100~days ($\Delta
m_{25-75}=0.5$~mags for $\eta_\mathrm{Ni}=0.5$, Figure~\ref{figure:etaNi_dm})
The common values of $\eta_\mathrm{Ni}$ for the observed SNe~IIP is
0.3\,--\,0.7 \citep{2016ApJ...823..127N}, therefore, \Ni{} \, significantly contributes to plateau shape and
duration.

Among other findings are:
\begin{itemize}
\item Regardless explosion energy and the total amount of radioactive \Ni{},
\Ni{} starts to affect plateau luminosity at particular time according to
degree of mixing. Particularly, centrally located \Ni{} modifies the light
curve at the end of the plateau, around day~75, while fully mixed \Ni{}
increases plateau luminosity at the most beginning, around day~20. In all cases 
this time depends only on the extent of \Ni{} distribution (see
Figure~\ref{figure:m12lbol}).
\item For the typical values of $\eta_\mathrm{Ni}$ (between 0.3 and 0.7), it is
difficult to distinguish contribution from pure recombination and cooling, and
\Ni{} heating. Light curves alone do not provide enough information to
differenciate between different degrees of mixing.
The observed SN\,2009ib with the long 150-day plateau has $\eta_\mathrm{Ni}=2.6$
and requires moderate amount of \Ni{} (0.046~\Msun{}) mixed heavily throughout the ejecta.
\item ``Boxcar'' mixing of \Ni{} leads to a bolometric light curve with a double step
transition from the plateau to the tail if explosion energy is relatively
low, about 0.5~foe. This kind of feature is not observed in normal SNe~IIP.
However, additional numerical simulations are required to clarify this
conclusion.
\item A centrally concentrated \Ni{} and ``boxcar'' lead to a sharp jump in \emph{B}--\emph{V} 
following the end of the plateau, which is not seen in observations. However, this aspect 
requires additional simulations accounting for non-LTE radiative transport.
\item There is no significant modification to photospheric velocity up to
day~50 due to \Ni{}-heating. However, moderate changes occur later.
\end{itemize}

We highlight that uniformly mixed \Ni{} either in half of the ejecta or
almost entire ejecta supports plateau luminosity providing light curves
consistent with observetions.
``Boxcar'' and centrally concentrated \Ni{} results in light curves which
colour evolution is inconsistent with observations. However, this should be
confirmed by additional numerical simulations.
If confirmed, this challenges the core-collapse explosion simulations, since
conventional ``boxcar'' \Ni{} mixing is believed to mimic the realistic macroscopic mixing 
in the supernova ejecta and which is predicted by core-collapse simulations 
\citep{2015A&A...577A..48W,2017MNRAS.472..491M,2017ApJ...846...37U}.

%% file: append.tex
\appendix
\section[The SN ejecta structure at coasting phase]{The SN ejecta structure at coasting phase}
\label{appendix:append}

\begin{figure}
\centering
\includegraphics[width=0.5\textwidth]{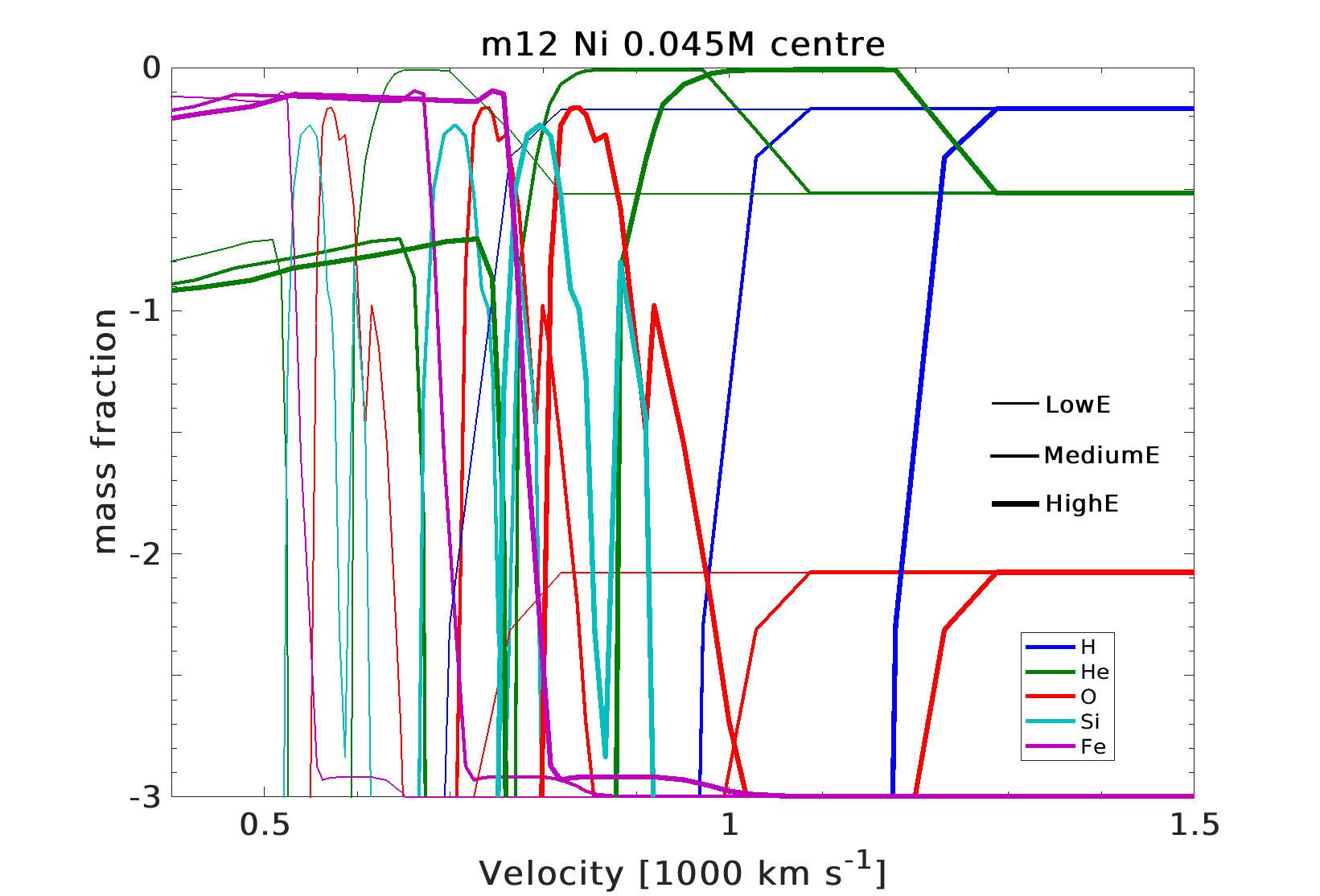}
\caption{Selected species, hydrogen (H), helium (He), oxygen (O), silicon
(Si), and iron (Fe), at day~170 for the model m12 with 0.045~\Msun{} of
\Ni{} distributed in the central part of the SN ejecta. ``LowE'' stands for
low explosion energy, i.e. 0.4~foe in our study, ``MediumE'' stands for
medium energy, i.e. 0.9~foe, and ``HighE'' stands for high energy, i.e.
1.35~foe.}
\label{figure:coast1}
\end{figure}

\begin{figure}
\centering
\includegraphics[width=0.5\textwidth]{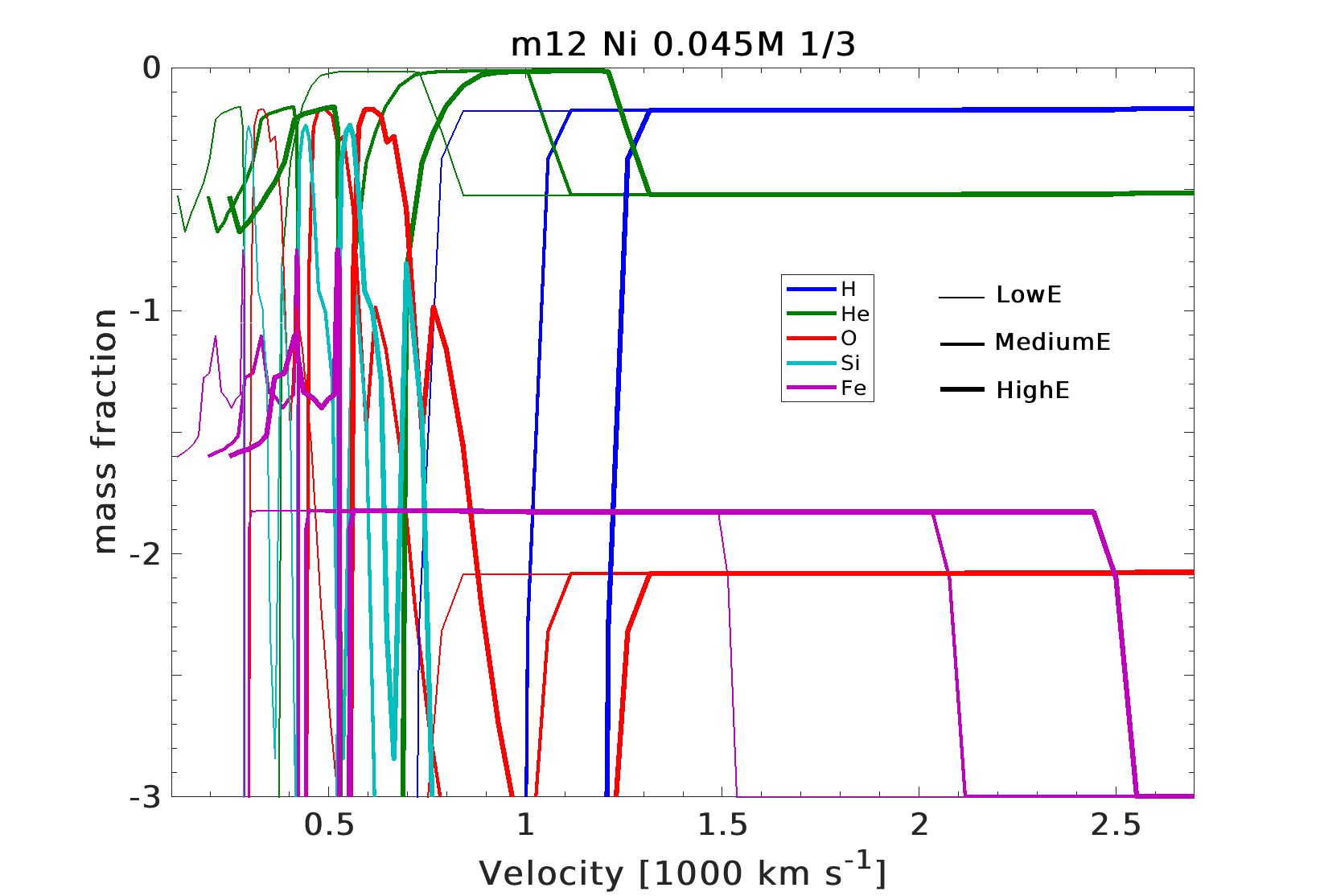}
\caption{The same as in Figure~\ref{figure:coast1} but for \Ni{} distributed
uniformly in 1/3 of the ejecta (case ``1/3'').}
\label{figure:coast2}
\end{figure}

\begin{figure}
\centering
\includegraphics[width=0.5\textwidth]{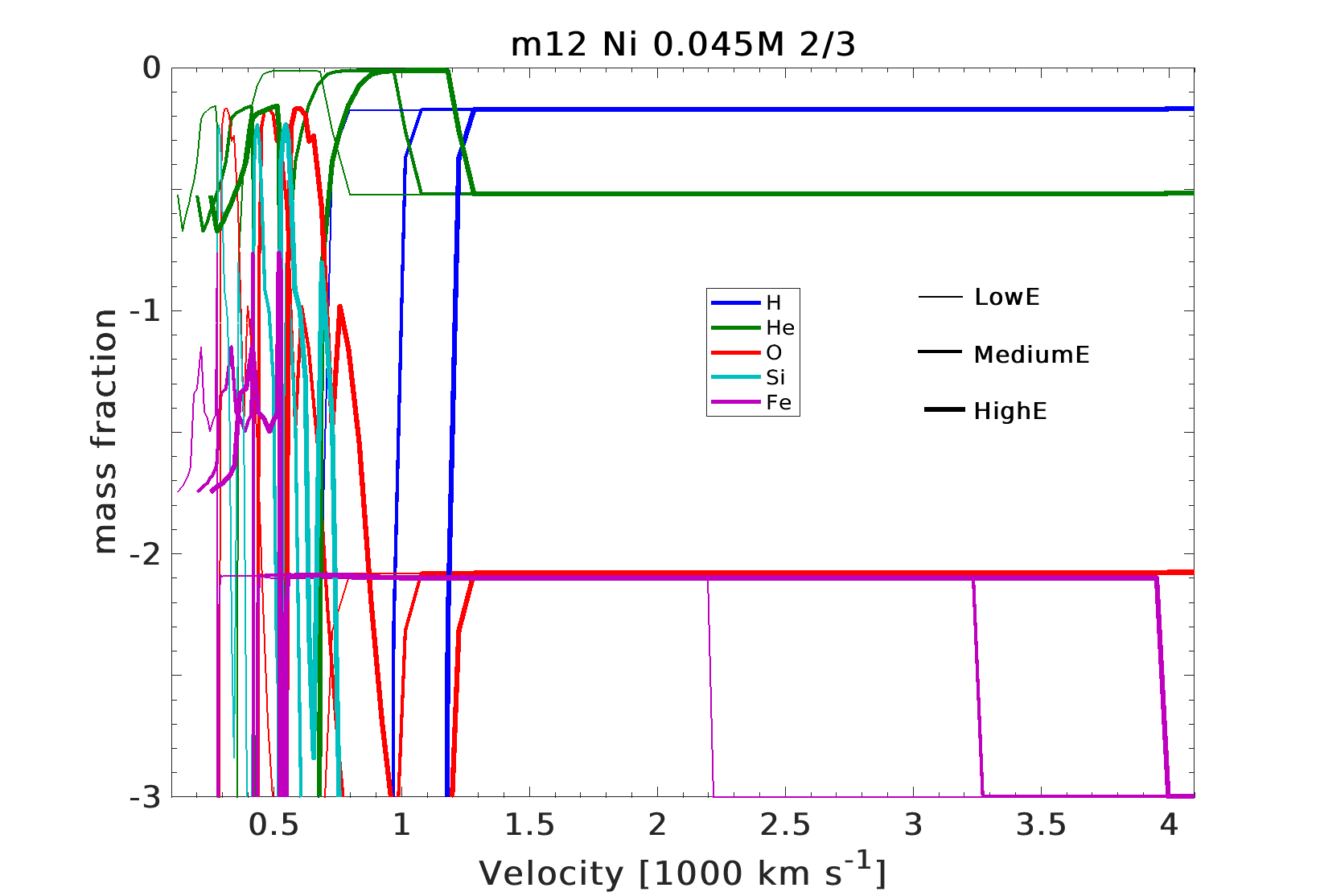}
\caption{The same as in Figure~\ref{figure:coast1} but for \Ni{} distributed
uniformly in 2/3 of the ejecta (case ``2/3'').}
\label{figure:coast3}
\end{figure}

\begin{figure}
\centering
\includegraphics[width=0.5\textwidth]{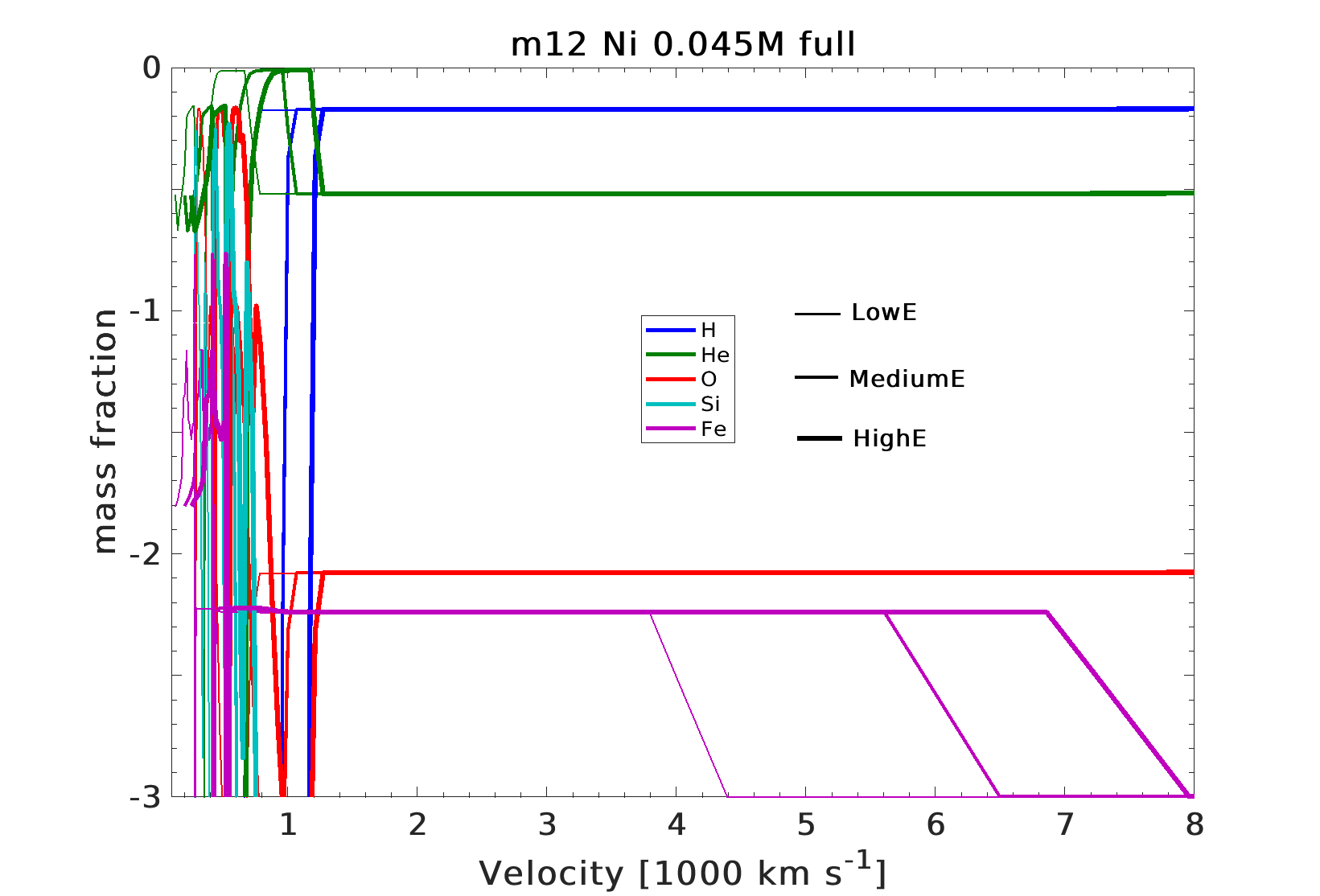}
\caption{The same as in Figure~\ref{figure:coast1} but for \Ni{} distributed
uniformly in entire ejecta (case ``full'').}
\label{figure:coast4}
\end{figure}

\begin{figure}
\centering
\includegraphics[width=0.5\textwidth]{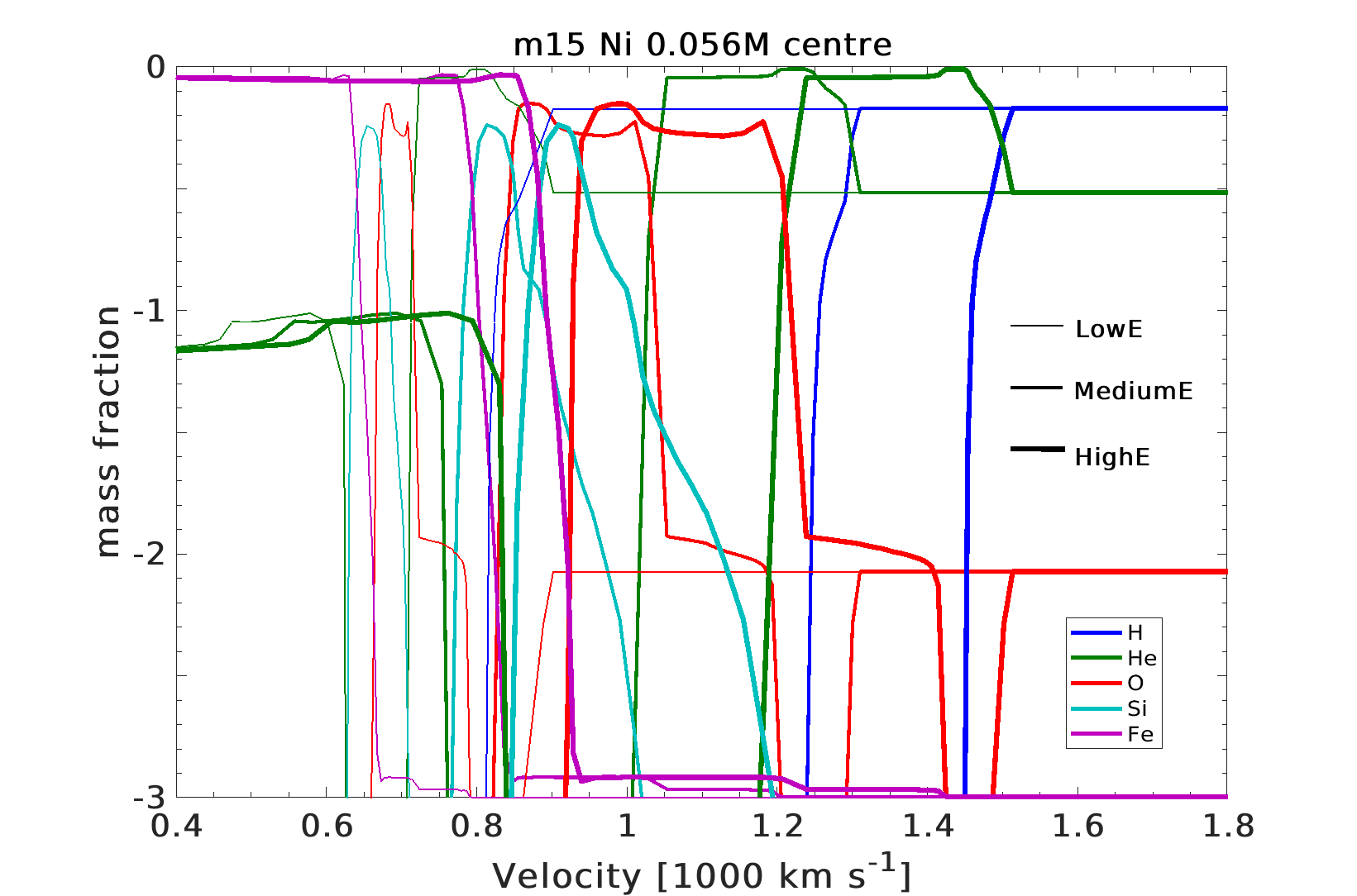}
\caption{The same as in Figure~\ref{figure:coast1} but for the model m15 with
0.056~\Msun{} of \Ni{} distributed in the central part of the SN ejecta. ``LowE'' stands for
low explosion energy, i.e. 0.53~foe in our study, ``MediumE'' stands for
medium energy, i.e. 1.1~foe, and ``HighE'' stands for high energy, i.e.
1.53~foe.}
\label{figure:coast5}
\end{figure}

\begin{figure}
\centering
\includegraphics[width=0.5\textwidth]{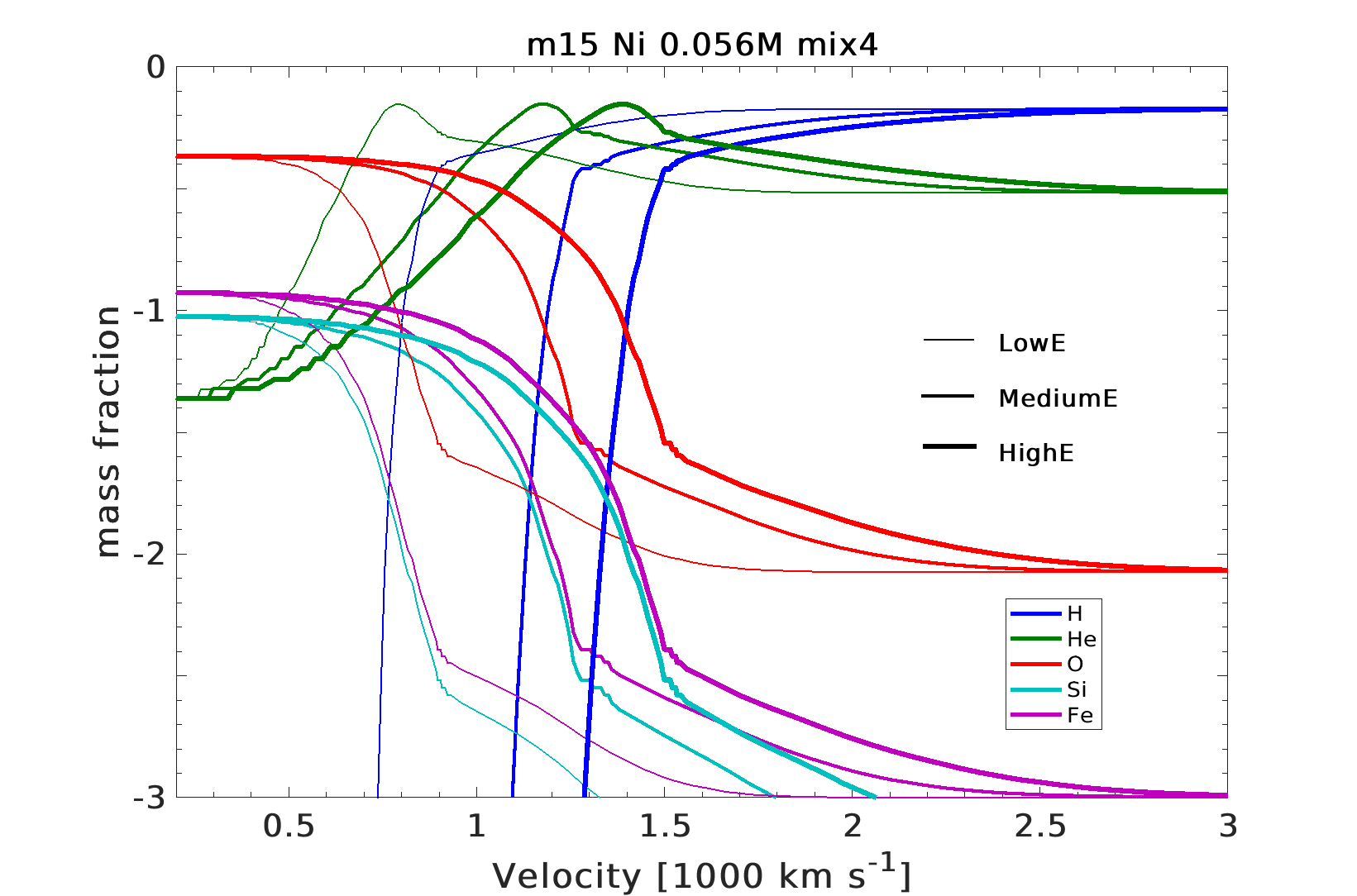}
\caption{The same as in Figure~\ref{figure:coast5} but for \Ni{} distributed
in a boxcar manner (case ``mix4'').}
\label{figure:coast6}
\end{figure}

\begin{figure}
\centering
\includegraphics[width=0.5\textwidth]{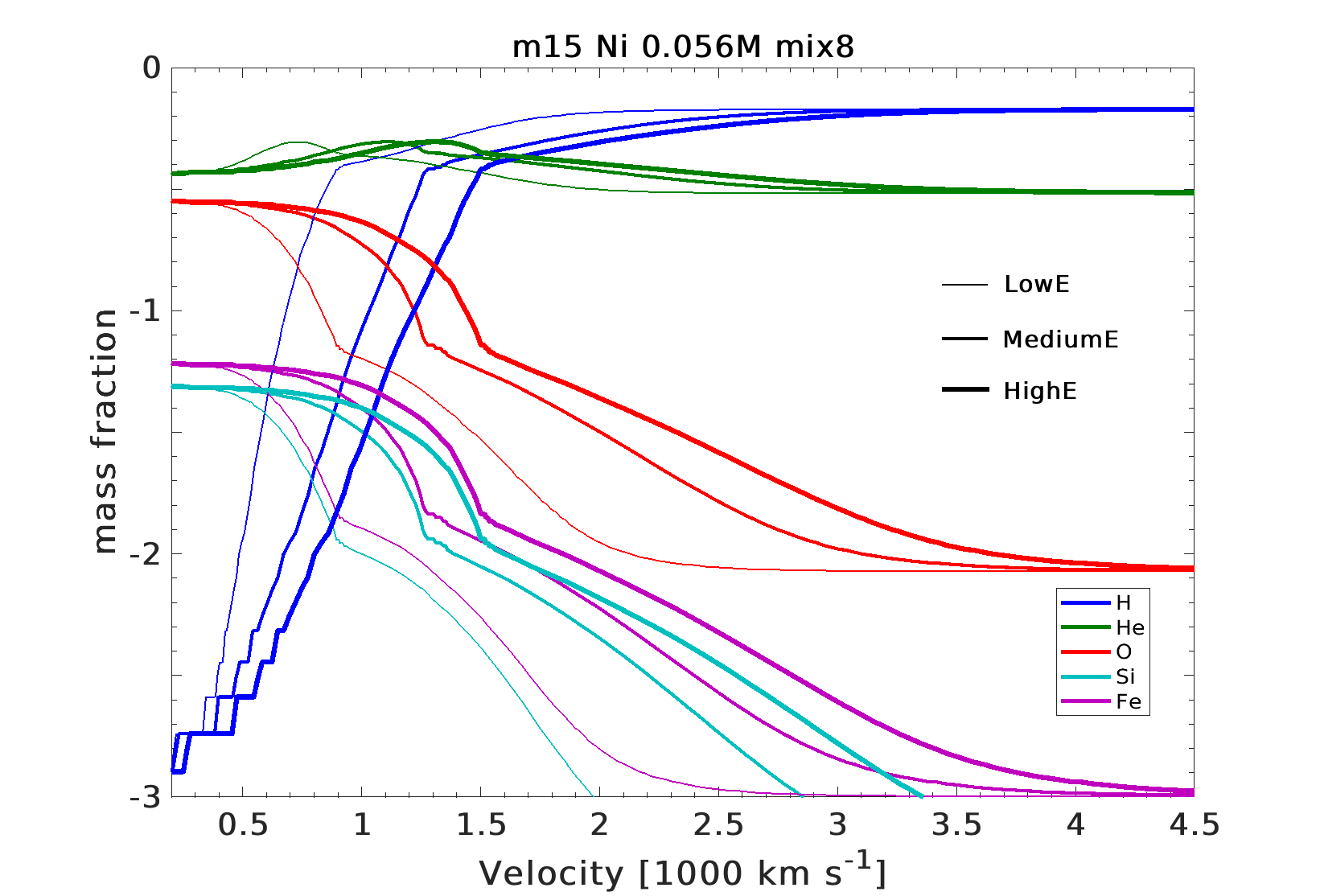}
\caption{The same as in Figure~\ref{figure:coast5} but for \Ni{} distributed
in a boxcar manner (case ``mix8'').}
\label{figure:coast7}
\end{figure}

\begin{figure}
\centering
\includegraphics[width=0.5\textwidth]{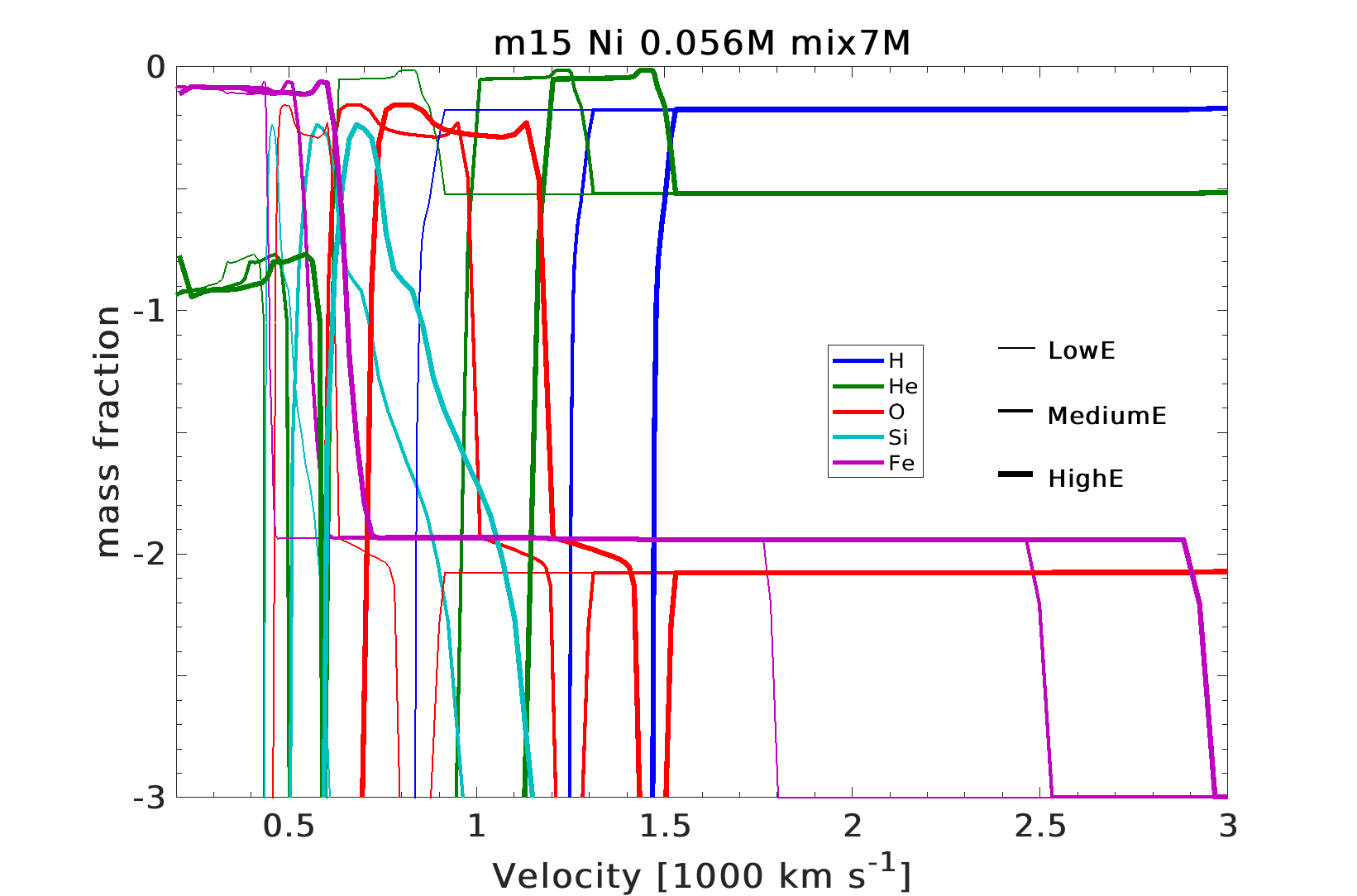}
\caption{The same as in Figure~\ref{figure:coast5} but for \Ni{} distributed
uniformly in half of the ejecta (inner 7~\Msun{}, case ``mix7M'').}
\label{figure:coast8}
\end{figure}

\begin{figure}
\centering
\includegraphics[width=0.5\textwidth]{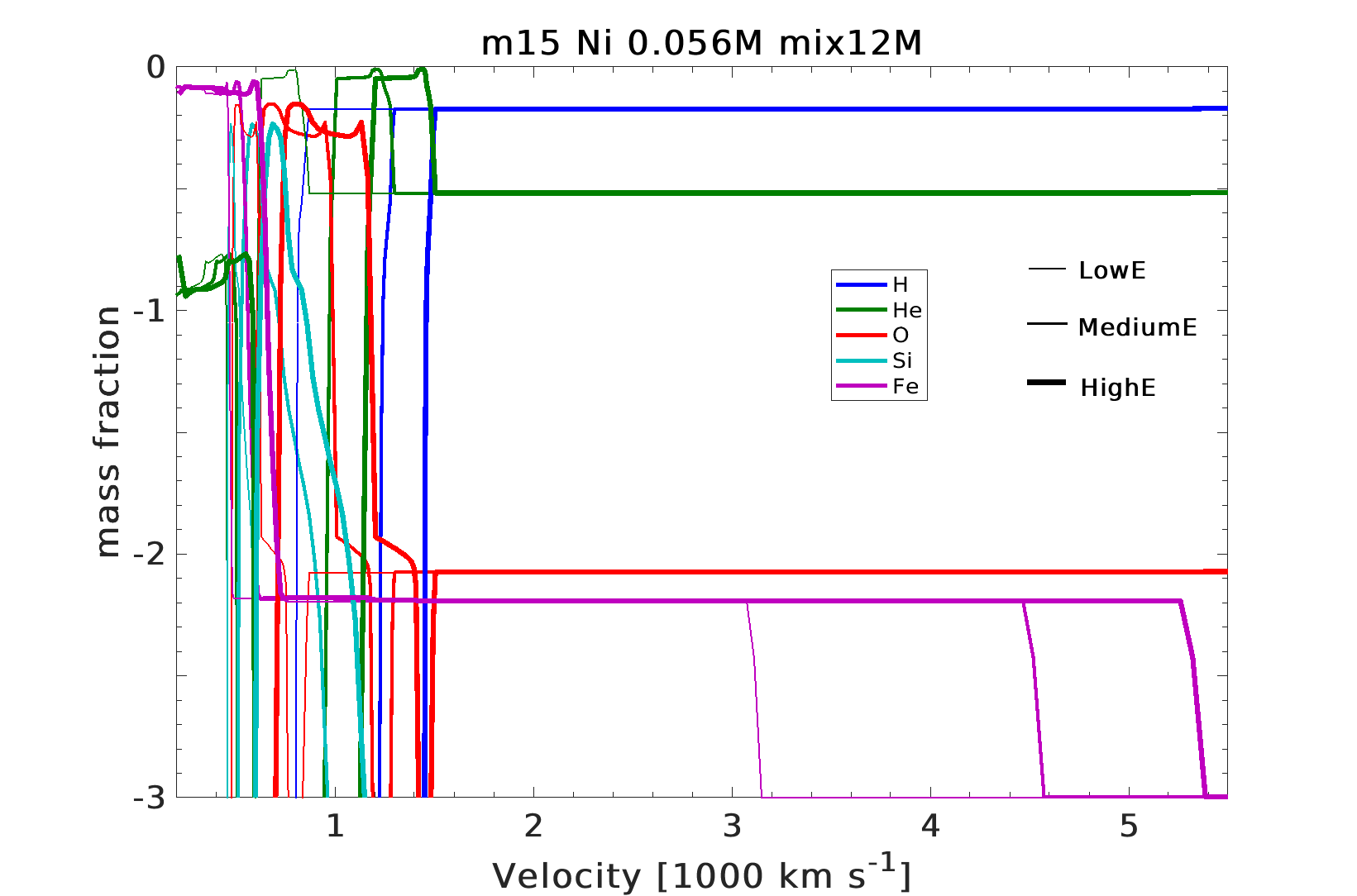}
\caption{The same as in Figure~\ref{figure:coast5} but for \Ni{} distributed
uniformly in entire ejecta (inner 12~\Msun{}, case ``mix12M'').}
\label{figure:coast9}
\end{figure}

Our study is focused on the light curve analysis, however, we present
additionally the SN ejecta structure at coasting phase. This might be
helpful for observers to interpret observational properties of a given
SN, like width of spectral lines of particular elements. In
Figures~\ref{figure:coast1}, \ref{figure:coast2}, \ref{figure:coast3}, and \ref{figure:coast4},
we show the selected species: hydrogen (H), helium (He), oxygen (O), silicon
(Si), and iron (Fe), at day~170 for the model m12 with 0.045~\Msun{} of
\Ni{} for all considered distributions of \Ni{} for this model and for all
cases of explosion energy, 0.4~foe, 0.9~foe, and 1.35~foe.
Similarly, Figures~\ref{figure:coast5}, \ref{figure:coast6}, \ref{figure:coast7}, 
\ref{figure:coast8}, and \ref{figure:coast9} present ejecta structure of the
model m15 with 0.056~\Msun{} of \Ni{} and considered explosion energies 
0.53~foe, 1.1~foe, and 1.53~foe.
Iron (``Fe'') in the Figures represent a sum of mass fractions of iron-group elements
included in the \verb|STELLA| simulations, i.e. iron, cobalt, and nickel.

Obviously, all chemical interfaces shift forward in velocity space for
higher energy. For instance, the outer boundary of iron-rich material moves at
1550~km\,s$^{\,-1}$ for the model m12 and the mixing case ``1/3'' exploded
with 0.4~foe (Figure~\ref{figure:coast2}), while it moves at 2550~km\,s$^{\,-1}$
for the explosion with 1.35~foe.
There is no big difference for distribution of all species except iron for different degree of uniform
mixing, since we limit our study and focus on modified \Ni{} distribution
(e.g. if compare Figures~\ref{figure:coast2} and \ref{figure:coast3}). 

\begin{figure}
\centering
\includegraphics[width=0.5\textwidth]{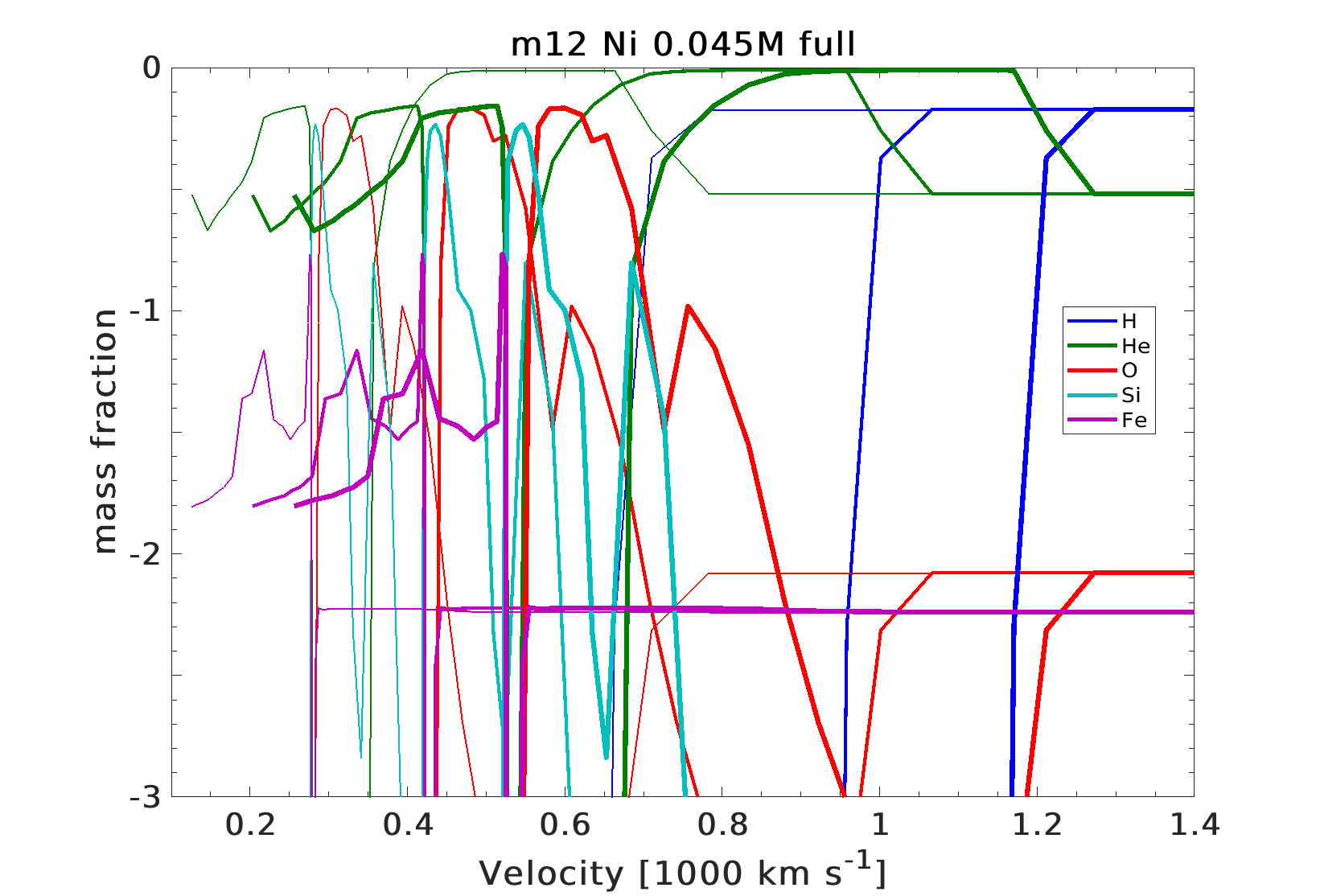}
\caption{The same as in Figure~\ref{figure:coast4} but for the inner part of
the ejecta.}
\label{figure:coast4a}
\end{figure}

\begin{figure}
\centering
\includegraphics[width=0.5\textwidth]{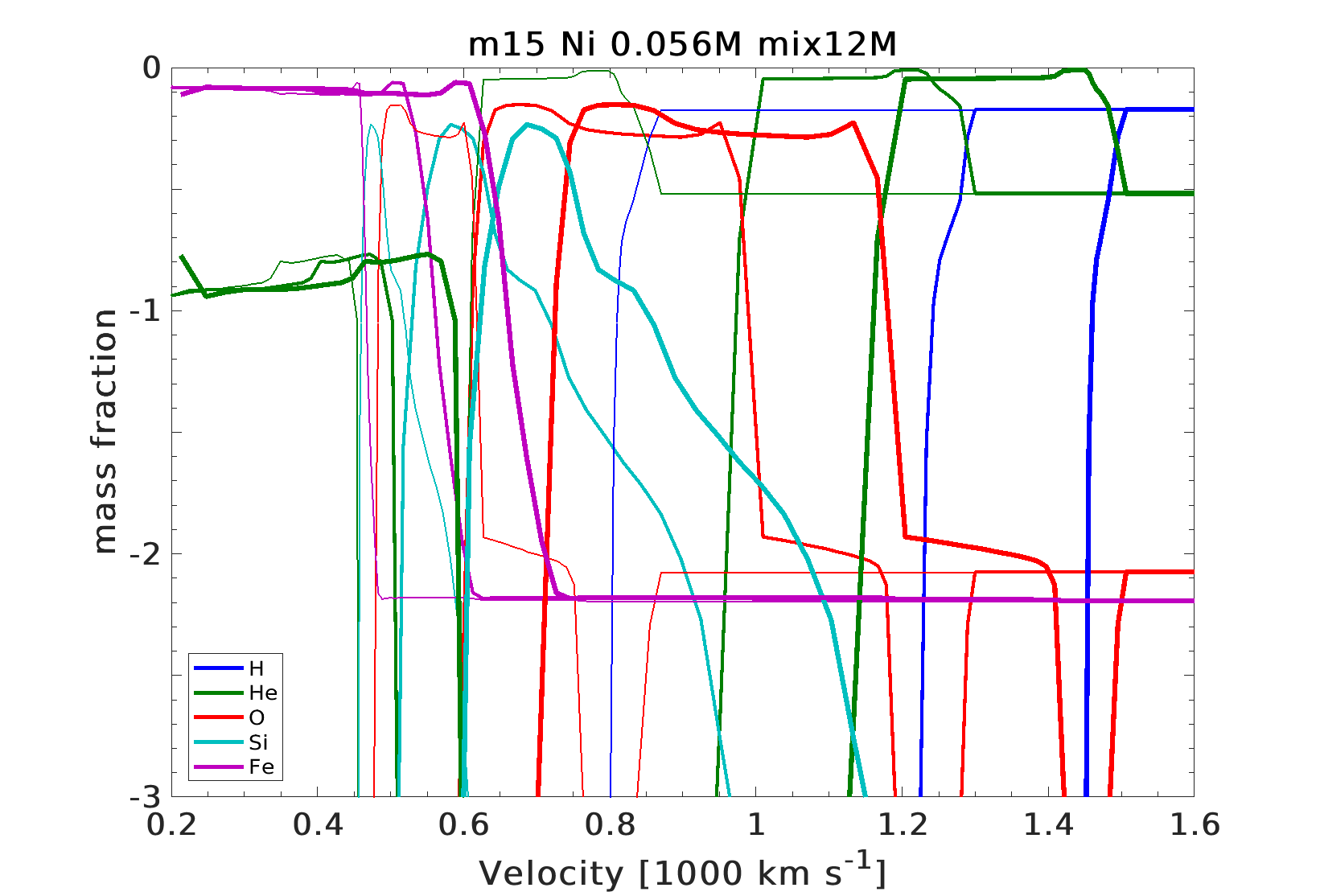}
\caption{The same as in Figure~\ref{figure:coast9} but for the inner part of
the ejecta.}
\label{figure:coast9a}
\end{figure}

%% file: platomix3.bbl
\begin{thebibliography}{}
\makeatletter
\relax
\def\mn@urlcharsother{\let\do\@makeother \do\$\do\&\do\#\do\^\do\_\do\%\do\~}
\def\mn@doi{\begingroup\mn@urlcharsother \@ifnextchar [ {\mn@doi@}
  {\mn@doi@[]}}
\def\mn@doi@[#1]#2{\def\@tempa{#1}\ifx\@tempa\@empty \href
  {http://dx.doi.org/#2} {doi:#2}\else \href {http://dx.doi.org/#2} {#1}\fi
  \endgroup}
\def\mn@eprint#1#2{\mn@eprint@#1:#2::\@nil}
\def\mn@eprint@arXiv#1{\href {http://arxiv.org/abs/#1} {{\tt arXiv:#1}}}
\def\mn@eprint@dblp#1{\href {http://dblp.uni-trier.de/rec/bibtex/#1.xml}
  {dblp:#1}}
\def\mn@eprint@#1:#2:#3:#4\@nil{\def\@tempa {#1}\def\@tempb {#2}\def\@tempc
  {#3}\ifx \@tempc \@empty \let \@tempc \@tempb \let \@tempb \@tempa \fi \ifx
  \@tempb \@empty \def\@tempb {arXiv}\fi \@ifundefined
  {mn@eprint@\@tempb}{\@tempb:\@tempc}{\expandafter \expandafter \csname
  mn@eprint@\@tempb\endcsname \expandafter{\@tempc}}}

\bibitem[\protect\citeauthoryear{{Anderson} et~al.,}{{Anderson}
  et~al.}{2014}]{2014ApJ...786...67A}
{Anderson} J.~P.,  et~al., 2014, \mn@doi [\apj] {10.1088/0004-637X/786/1/67},
  \href {http://adsabs.harvard.edu/abs/2014ApJ...786...67A} {786, 67}

\bibitem[\protect\citeauthoryear{{Bersten}, {Benvenuto}  \& {Hamuy}}{{Bersten}
  et~al.}{2011}]{2011ApJ...729...61B}
{Bersten} M.~C.,  {Benvenuto} O.,   {Hamuy} M.,  2011, \mn@doi [\apj]
  {10.1088/0004-637X/729/1/61}, \href
  {http://adsabs.harvard.edu/abs/2011ApJ...729...61B} {729, 61}

\bibitem[\protect\citeauthoryear{{Blinnikov}, {Eastman}, {Bartunov},
  {Popolitov}  \& {Woosley}}{{Blinnikov} et~al.}{1998}]{1998ApJ...496..454B}
{Blinnikov} S.~I.,  {Eastman} R.,  {Bartunov} O.~S.,  {Popolitov} V.~A.,
  {Woosley} S.~E.,  1998, \mn@doi [\apj] {10.1086/305375}, \href
  {http://adsabs.harvard.edu/abs/1998ApJ...496..454B} {496, 454}

\bibitem[\protect\citeauthoryear{{Blinnikov}, {R{\"o}pke}, {Sorokina},
  {Gieseler}, {Reinecke}, {Travaglio}, {Hillebrandt}  \&
  {Stritzinger}}{{Blinnikov} et~al.}{2006}]{2006A&A...453..229B}
{Blinnikov} S.~I.,  {R{\"o}pke} F.~K.,  {Sorokina} E.~I.,  {Gieseler} M.,
  {Reinecke} M.,  {Travaglio} C.,  {Hillebrandt} W.,   {Stritzinger} M.,  2006,
  \mn@doi [\aap] {10.1051/0004-6361:20054594}, \href
  {http://adsabs.harvard.edu/abs/2006A%26A...453..229B} {453, 229}

\bibitem[\protect\citeauthoryear{{Chugai}}{{Chugai}}{1991}]{1991SvAL...17..210C}
{Chugai} N.~N.,  1991, Soviet Astronomy Letters, \href
  {http://adsabs.harvard.edu/abs/1991SvAL...17..210C} {17, 210}

\bibitem[\protect\citeauthoryear{{Dahlen} et~al.,}{{Dahlen}
  et~al.}{2004}]{2004ApJ...613..189D}
{Dahlen} T.,  et~al., 2004, \mn@doi [\apj] {10.1086/422899}, \href
  {http://adsabs.harvard.edu/abs/2004ApJ...613..189D} {613, 189}

\bibitem[\protect\citeauthoryear{{Dessart} \& {Hillier}}{{Dessart} \&
  {Hillier}}{2010}]{2010MNRAS.405.2141D}
{Dessart} L.,  {Hillier} D.~J.,  2010, \mn@doi [\mnras]
  {10.1111/j.1365-2966.2010.16611.x}, \href
  {http://adsabs.harvard.edu/abs/2010MNRAS.405.2141D} {405, 2141}

\bibitem[\protect\citeauthoryear{{Dessart} \& {Hillier}}{{Dessart} \&
  {Hillier}}{2011}]{2011MNRAS.410.1739D}
{Dessart} L.,  {Hillier} D.~J.,  2011, \mn@doi [\mnras]
  {10.1111/j.1365-2966.2010.17557.x}, \href
  {http://adsabs.harvard.edu/abs/2011MNRAS.410.1739D} {410, 1739}

\bibitem[\protect\citeauthoryear{{Dessart}, {Livne}  \& {Waldman}}{{Dessart}
  et~al.}{2010}]{2010MNRAS.405.2113D}
{Dessart} L.,  {Livne} E.,   {Waldman} R.,  2010, \mn@doi [\mnras]
  {10.1111/j.1365-2966.2010.16626.x}, \href
  {http://adsabs.harvard.edu/abs/2010MNRAS.405.2113D} {405, 2113}

\bibitem[\protect\citeauthoryear{{Dessart}, {Audit}  \& {Hillier}}{{Dessart}
  et~al.}{2015}]{2015MNRAS.449.4304D}
{Dessart} L.,  {Audit} E.,   {Hillier} D.~J.,  2015, \mn@doi [\mnras]
  {10.1093/mnras/stv609}, \href
  {http://adsabs.harvard.edu/abs/2015MNRAS.449.4304D} {449, 4304}

\bibitem[\protect\citeauthoryear{{Falk} \& {Arnett}}{{Falk} \&
  {Arnett}}{1977}]{1977ApJS...33..515F}
{Falk} S.~W.,  {Arnett} W.~D.,  1977, \mn@doi [\apjs] {10.1086/190440}, \href
  {http://adsabs.harvard.edu/abs/1977ApJS...33..515F} {33, 515}

\bibitem[\protect\citeauthoryear{{Faran} et~al.,}{{Faran}
  et~al.}{2014a}]{2014MNRAS.442..844F}
{Faran} T.,  et~al., 2014a, \mn@doi [\mnras] {10.1093/mnras/stu955}, \href
  {http://adsabs.harvard.edu/abs/2014MNRAS.442..844F} {442, 844}

\bibitem[\protect\citeauthoryear{{Faran} et~al.,}{{Faran}
  et~al.}{2014b}]{2014MNRAS.445..554F}
{Faran} T.,  et~al., 2014b, \mn@doi [\mnras] {10.1093/mnras/stu1760}, \href
  {http://adsabs.harvard.edu/abs/2014MNRAS.445..554F} {445, 554}

\bibitem[\protect\citeauthoryear{{Grasberg} \& {Nadezhin}}{{Grasberg} \&
  {Nadezhin}}{1976}]{1976Ap&SS..44..409G}
{Grasberg} E.~K.,  {Nadezhin} D.~K.,  1976, \mn@doi [\apss]
  {10.1007/BF00642529}, \href
  {http://adsabs.harvard.edu/abs/1976Ap%26SS..44..409G} {44, 409}

\bibitem[\protect\citeauthoryear{{Grasberg}, {Imshenik}  \&
  {Nadyozhin}}{{Grasberg} et~al.}{1971}]{1971Ap&SS..10....3G}
{Grasberg} E.~K.,  {Imshenik} V.~S.,   {Nadyozhin} D.~K.,  1971, \mn@doi
  [\apss] {10.1007/BF00654603}, \href
  {http://adsabs.harvard.edu/abs/1971Ap%26SS..10....3G} {10, 3}

\bibitem[\protect\citeauthoryear{{Kasen} \& {Woosley}}{{Kasen} \&
  {Woosley}}{2009}]{2009ApJ...703.2205K}
{Kasen} D.,  {Woosley} S.~E.,  2009, \mn@doi [\apj]
  {10.1088/0004-637X/703/2/2205}, \href
  {http://adsabs.harvard.edu/abs/2009ApJ...703.2205K} {703, 2205}

\bibitem[\protect\citeauthoryear{{Katz}, {Kushnir}  \& {Dong}}{{Katz}
  et~al.}{2013}]{2013arXiv1301.6766K}
{Katz} B.,  {Kushnir} D.,   {Dong} S.,  2013, preprint, \href
  {http://adsabs.harvard.edu/abs/2013arXiv1301.6766K} {} (\mn@eprint {arXiv}
  {1301.6766})

\bibitem[\protect\citeauthoryear{{Li}, {Chornock}, {Leaman}, {Filippenko},
  {Poznanski}, {Wang}, {Ganeshalingam}  \& {Mannucci}}{{Li}
  et~al.}{2011}]{2011MNRAS.412.1473L}
{Li} W.,  {Chornock} R.,  {Leaman} J.,  {Filippenko} A.~V.,  {Poznanski} D.,
  {Wang} X.,  {Ganeshalingam} M.,   {Mannucci} F.,  2011, \mn@doi [\mnras]
  {10.1111/j.1365-2966.2011.18162.x}, \href
  {http://adsabs.harvard.edu/abs/2011MNRAS.412.1473L} {412, 1473}

\bibitem[\protect\citeauthoryear{{Li}, {Hillier}  \& {Dessart}}{{Li}
  et~al.}{2012}]{2012MNRAS.426.1671L}
{Li} C.,  {Hillier} D.~J.,   {Dessart} L.,  2012, \mn@doi [\mnras]
  {10.1111/j.1365-2966.2012.21198.x}, \href
  {http://adsabs.harvard.edu/abs/2012MNRAS.426.1671L} {426, 1671}

\bibitem[\protect\citeauthoryear{{Litvinova} \& {Nadezhin}}{{Litvinova} \&
  {Nadezhin}}{1985}]{1985SvAL...11..145L}
{Litvinova} I.~Y.,  {Nadezhin} D.~K.,  1985, Soviet Astronomy Letters, \href
  {http://adsabs.harvard.edu/abs/1985SvAL...11..145L} {11, 145}

\bibitem[\protect\citeauthoryear{{Livne}}{{Livne}}{1993}]{1993ApJ...412..634L}
{Livne} E.,  1993, \mn@doi [\apj] {10.1086/172950}, \href
  {http://adsabs.harvard.edu/abs/1993ApJ...412..634L} {412, 634}

\bibitem[\protect\citeauthoryear{{Mackey}, {Bromm}  \& {Hernquist}}{{Mackey}
  et~al.}{2003}]{2003ApJ...586....1M}
{Mackey} J.,  {Bromm} V.,   {Hernquist} L.,  2003, \mn@doi [\apj]
  {10.1086/367613}, \href {http://adsabs.harvard.edu/abs/2003ApJ...586....1M}
  {586, 1}

\bibitem[\protect\citeauthoryear{{Mannucci}, {Della Valle}  \&
  {Panagia}}{{Mannucci} et~al.}{2007}]{2007MNRAS.377.1229M}
{Mannucci} F.,  {Della Valle} M.,   {Panagia} N.,  2007, \mn@doi [\mnras]
  {10.1111/j.1365-2966.2007.11676.x}, \href
  {http://adsabs.harvard.edu/abs/2007MNRAS.377.1229M} {377, 1229}

\bibitem[\protect\citeauthoryear{{M{\"u}ller}, {Melson}, {Heger}  \&
  {Janka}}{{M{\"u}ller} et~al.}{2017}]{2017MNRAS.472..491M}
{M{\"u}ller} B.,  {Melson} T.,  {Heger} A.,   {Janka} H.-T.,  2017, \mn@doi
  [\mnras] {10.1093/mnras/stx1962}, \href
  {http://adsabs.harvard.edu/abs/2017MNRAS.472..491M} {472, 491}

\bibitem[\protect\citeauthoryear{{Nakar}, {Poznanski}  \& {Katz}}{{Nakar}
  et~al.}{2016}]{2016ApJ...823..127N}
{Nakar} E.,  {Poznanski} D.,   {Katz} B.,  2016, \mn@doi [\apj]
  {10.3847/0004-637X/823/2/127}, \href
  {http://adsabs.harvard.edu/abs/2016ApJ...823..127N} {823, 127}

\bibitem[\protect\citeauthoryear{{Paxton}, {Bildsten}, {Dotter}, {Herwig},
  {Lesaffre}  \& {Timmes}}{{Paxton} et~al.}{2011}]{2011ApJS..192....3P}
{Paxton} B.,  {Bildsten} L.,  {Dotter} A.,  {Herwig} F.,  {Lesaffre} P.,
  {Timmes} F.,  2011, \mn@doi [\apjs] {10.1088/0067-0049/192/1/3}, \href
  {http://adsabs.harvard.edu/abs/2011ApJS..192....3P} {192, 3}

\bibitem[\protect\citeauthoryear{{Paxton} et~al.,}{{Paxton}
  et~al.}{2013}]{2013ApJS..208....4P}
{Paxton} B.,  et~al., 2013, \mn@doi [\apjs] {10.1088/0067-0049/208/1/4}, \href
  {http://adsabs.harvard.edu/abs/2013ApJS..208....4P} {208, 4}

\bibitem[\protect\citeauthoryear{{Paxton} et~al.,}{{Paxton}
  et~al.}{2015}]{2015ApJS..220...15P}
{Paxton} B.,  et~al., 2015, \mn@doi [\apjs] {10.1088/0067-0049/220/1/15}, \href
  {http://adsabs.harvard.edu/abs/2015ApJS..220...15P} {220, 15}

\bibitem[\protect\citeauthoryear{{Popov}}{{Popov}}{1993}]{1993ApJ...414..712P}
{Popov} D.~V.,  1993, \mn@doi [\apj] {10.1086/173117}, \href
  {http://adsabs.harvard.edu/abs/1993ApJ...414..712P} {414, 712}

\bibitem[\protect\citeauthoryear{{Shklovskii}}{{Shklovskii}}{1960}]{1960SvA.....4..355S}
{Shklovskii} I.~S.,  1960, \sovast, \href
  {http://adsabs.harvard.edu/abs/1960SvA.....4..355S} {4, 355}

\bibitem[\protect\citeauthoryear{{Shussman}, {Nakar}, {Waldman}  \&
  {Katz}}{{Shussman} et~al.}{2016}]{2016arXiv160202774S}
{Shussman} T.,  {Nakar} E.,  {Waldman} R.,   {Katz} B.,  2016, preprint, \href
  {http://adsabs.harvard.edu/abs/2016arXiv160202774S} {} (\mn@eprint {arXiv}
  {1602.02774})

\bibitem[\protect\citeauthoryear{{Smartt}}{{Smartt}}{2009}]{2009ARA&A..47...63S}
{Smartt} S.~J.,  2009, \mn@doi [\araa] {10.1146/annurev-astro-082708-101737},
  \href {http://adsabs.harvard.edu/abs/2009ARA%26A..47...63S} {47, 63}

\bibitem[\protect\citeauthoryear{{Smith}, {Li}, {Filippenko}  \&
  {Chornock}}{{Smith} et~al.}{2011}]{2011MNRAS.412.1522S}
{Smith} N.,  {Li} W.,  {Filippenko} A.~V.,   {Chornock} R.,  2011, \mn@doi
  [\mnras] {10.1111/j.1365-2966.2011.17229.x}, \href
  {http://adsabs.harvard.edu/abs/2011MNRAS.412.1522S} {412, 1522}

\bibitem[\protect\citeauthoryear{{Sukhbold}, {Ertl}, {Woosley}, {Brown}  \&
  {Janka}}{{Sukhbold} et~al.}{2016}]{2016ApJ...821...38S}
{Sukhbold} T.,  {Ertl} T.,  {Woosley} S.~E.,  {Brown} J.~M.,   {Janka} H.-T.,
  2016, \mn@doi [\apj] {10.3847/0004-637X/821/1/38}, \href
  {http://adsabs.harvard.edu/abs/2016ApJ...821...38S} {821, 38}

\bibitem[\protect\citeauthoryear{{Tak{\'a}ts} et~al.,}{{Tak{\'a}ts}
  et~al.}{2015}]{2015MNRAS.450.3137T}
{Tak{\'a}ts} K.,  et~al., 2015, \mn@doi [\mnras] {10.1093/mnras/stv857}, \href
  {http://adsabs.harvard.edu/abs/2015MNRAS.450.3137T} {450, 3137}

\bibitem[\protect\citeauthoryear{{Utrobin}, {Wongwathanarat}, {Janka}  \&
  {M{\"u}ller}}{{Utrobin} et~al.}{2017}]{2017ApJ...846...37U}
{Utrobin} V.~P.,  {Wongwathanarat} A.,  {Janka} H.-T.,   {M{\"u}ller} E.,
  2017, \mn@doi [\apj] {10.3847/1538-4357/aa8594}, \href
  {http://adsabs.harvard.edu/abs/2017ApJ...846...37U} {846, 37}

\bibitem[\protect\citeauthoryear{{Valenti} et~al.,}{{Valenti}
  et~al.}{2016}]{2016MNRAS.459.3939V}
{Valenti} S.,  et~al., 2016, \mn@doi [\mnras] {10.1093/mnras/stw870}, \href
  {http://adsabs.harvard.edu/abs/2016MNRAS.459.3939V} {459, 3939}

\bibitem[\protect\citeauthoryear{{Wongwathanarat}, {M{\"u}ller}  \&
  {Janka}}{{Wongwathanarat} et~al.}{2015}]{2015A&A...577A..48W}
{Wongwathanarat} A.,  {M{\"u}ller} E.,   {Janka} H.-T.,  2015, \mn@doi [\aap]
  {10.1051/0004-6361/201425025}, \href
  {http://adsabs.harvard.edu/abs/2015A%26A...577A..48W} {577, A48}

\bibitem[\protect\citeauthoryear{{Young}}{{Young}}{2004}]{2004ApJ...617.1233Y}
{Young} T.~R.,  2004, \mn@doi [\apj] {10.1086/425675}, \href
  {http://adsabs.harvard.edu/abs/2004ApJ...617.1233Y} {617, 1233}

\makeatother
\end{thebibliography}
